\documentclass[prl,aps,twocolumn,groupedaddress,floats,final,superscriptaddress,nopacks]{revtex4-1}
\usepackage{graphicx}
\usepackage{subfigure}
\usepackage{amssymb}
\usepackage{latexsym}
\usepackage{epsfig}
\usepackage{psfrag}
\usepackage{dcolumn}
\usepackage{amsmath,amsfonts,amssymb,bm,ulem}
\usepackage{color}
\usepackage{float}

\newcommand{\be}{\begin{equation}}
\newcommand{\ee}{\end{equation}}
\newcommand{\bea}{\begin{eqnarray}}
\newcommand{\eea}{\end{eqnarray}}

\begin{document}
\title{Real-frequency response functions at finite temperature}

\author{I. S. Tupitsyn}
\affiliation{Department of Physics, University of Massachusetts, Amherst, MA 01003, USA}
\author{A. M. Tsvelik}
\affiliation{Condensed Matter Physics and Materials Science Division, Brookhaven National Laboratory, Upton, NY 11973-5000, USA}
\author{R. M. Konik}
\affiliation{Condensed Matter Physics and Materials Science Division, Brookhaven National Laboratory, Upton, NY 11973-5000, USA}
\author{N. V. Prokof'ev}
\affiliation{Department of Physics, University of Massachusetts, Amherst, MA 01003, USA}


\begin{abstract}
Building on previous developments \cite{LeBlanc2019,Holm98,ferreroT2}, we show that the Diagrammatic Monte Carlo technique allows to compute finite temperature response functions directly on the real-frequency axis within any field-theoretical formulation of the interacting fermion problem. There are no limitations on the type and nature of the system's action or whether partial summation and self-consistent treatment of certain diagram classes are used. In particular, by eliminating the need for numerical analytic continuation from a Matsubara representation, our scheme allows to study
spectral densities of arbitrary complexity with controlled accuracy in models with frequency-dependent effective interactions. For illustrative purposes we consider the problem of the plasmon line-width
in a homogeneous electron gas (jellium).
\end{abstract}

\maketitle


{\it Introduction.}  The promise of the Diagrammatic Monte Carlo (DiagMC) technique--stochastic sampling of high-order connected Feynman diagrams with extrapolation to the infinite diagram order limit--in solving the computational complexity problem for interacting fermions \cite{signproblem} ultimately rests on our ability to formulate a field-theoretical approach with a quickly converging series expansion. As is often the case in the strongly correlated regime, an expansion based on the original interaction potentials, $V_0$, and ``bare" fermion propagators, $G_0$, does not converge.
To proceed, the problem is transformed identically by incorporating certain classes of diagrams and interaction effects into an alternative ``starting point". This introduces new effective propagators, $\widetilde{G}$, interactions, $U$, and counter terms, $\Lambda$, in terms of which an alternative  diagrammatic expansion is formulated. Shifted and homotopic action tools \cite{ShiftAct,homotopic} allow to achieve this goal generically by expressing final answers as Taylor series in powers of the auxiliary parameter $\xi$, with $\xi=1$ corresponding to the original problem, and ensuring that the resulting series converge for any $\xi<1$.

To connect finite-temperature calculations with experimental probes not based on thermodynamic potentials, one needs to compute response functions at real frequencies, or spectral densities. The notorious problem faced by simulations performed in the Matsubara representation is a need for a numerical analytic continuation (NAC) procedure from the imaginary to the real-frequency domain. In general, NAC is only meaningful conditionally (by imposing constraints on the answer), and even extraordinary accurate Monte Carlo (MC) data cannot help resolve fine spectral features following broad lower-frequency peaks, or narrow Drude peaks in optical conductivity \cite{olya}. Until recently, the infamous NAC problem standing on the way of the accurate theoretical description of experimentally relevant observables was considered unavoidable.

The breakthrough development in the context of the DiagMC technique was reported in Ref.~\cite{LeBlanc2019} for the Hubbard model. The key observation was that for expansions in terms of $G_0$ and $V_0$, the summation over all internal fermionic Matsubara frequencies, $ \omega_n = 2\pi T (n+1/2)$ with integer $n$, can be performed analytically with the help of the Cauchy formula
\begin{equation}
T \sum_n \left( \prod_{j=1}^{M} \frac{1}{i\omega_n -a_j}  \right) = \sum_{j=1}^{M} n_j
\left( \prod_{s\ne j}^{M} \frac{1}{a_j-a_s}  \right) \, ,
\label{Cauchy}
\end{equation}
and automated for arbitrary high-order diagrams.
Here $n_j \equiv n(a_j)=[e^{a_j/T}+1]^{-1}$ is the Fermi-Dirac function. Bose-Einstein function is related to $n(a)$ by $N(a) = -n(a + i \pi T)$. Thus, summation over bosonic frequency $\omega_m$ is included in Eq.~(\ref{Cauchy}) by the transformation $i \omega_m - a  =   i \omega_{m+1/2} - a'$ with $a'=a + i \pi T$. [For the imaginary time implementation see Refs.~\cite{ferreroT,ferreroT2}.] The Wick rotation of external frequency from the imaginary to the real axis is then performed analytically by replacing $i\Omega_s$ with $\Omega +i\eta $, where $\eta \to 0$ is positive. The remaining integrals/sums are sampled by Monte Carlo to compute spectral densities directly without any need for NAC. [The proposed below scheme allows to take the $\eta \to 0$ limit analytically, see Supplemental material \cite{SM}.]

The entire procedure relies on (i) the simple pole structure of the bare Green's function,
$G_0=(i\omega_n - \epsilon_{\mathbf k}+\mu )^{-1}$, where $\epsilon_{\mathbf k}$ is the bare dispersion relation and $\mu$ is the chemical potential (we suppress the spin index for brevity), and (ii) a frequency independent interaction potential $V_0$. These requirements are not satisfied when the diagrammatic expansion is performed in terms of dressed/renormalized propagators and retarded effective interactions to produce convergent series in the strongly correlated regimes. Even if $\widetilde{G}$ and $U$ have transparent analytical structure in the Matsubara representation, the summation over all internal Matsubara frequencies cannot be performed analytically any more. For example, in the random phase approximation (RPA) for the homogeneous electron gas (jellium), the polarization operator in the effective screened interaction, $U^{-1}=V_0^{-1} - \Pi$, is approximated by the finite-temperature version of the Lindhard function \cite{Lindhard}. No diagram with these $U$-lines can be summed over Matsubara frequencies analytically.

It appears that conditions for performing real-frequency simulations are incompatible with the generic tools needed to obtain convergent series expansions. In this work we present a simple solution to this dilemma and formulate an approach that allows to compute real-frequency response functions within an arbitrary field-theoretical setup. To demonstrate how our approach works in practice, we compute the plasmon line-width, $\gamma_{pl}$, in the jellium model as a function of momentum and temperature. The problem of plasmon decay is under active study because of its importance for optoelectronics, photovoltaics, photocatalysis, and other applications (see, for instance, Refs.~\cite{Luther2013,Clavero2014,Atwater2010,Linic2011,Mukherjee2013,Li2015,Kolwas2019} and literature therein). Contrary to solid state materials where inter-band transitions and Umklapp processes are possible (and thus the plasmon line-width can be obtained within the lowest skeleton order diagrams, in the so-called GW approximation - see, for instance, Refs.~\cite{Bernardi2015,Sundar2014}), we find that meaningful results for $\gamma_{pl}$ in jellium crucially depend on vertex corrections.


{\it Real-frequency finite-temperature technique.} To perform the Wick rotation by the substitution $i\Omega_s \to \Omega +i \eta$, the function in question has to be known analytically. The key observation leading to solution is that at any point in the DiagMC simulation, the propagators and interactions used to express the diagram's contribution are assumed to be {\it known}, either analytically or numerically (from relatively simple auxiliary simulations). The first step is to convert this knowledge into spectral densities and use them to express $\tilde{G}$ and $U$ via
\bea
\tilde{G}(\mathbf{k},i\omega_n) &=&
\frac{1}{\pi} \int_{-\infty}^{\infty} du \frac{A(\mathbf{k},u)}{i\omega_n - u};
\label{Gspectral} \\
U(\mathbf{k},i\omega_s) &=&
V_0(k) + \frac{1}{\pi} \int_{-\infty}^{\infty} dv \frac{D(\mathbf{k},v)}{i\omega_s - v},
\label{Uspectral}
\eea
with bosonic Matsubara frequencies $\omega_s = 2\pi T s$. The second step is to rewrite all diagrammatic contributions in terms of the $A$ and $D$ functions. This will add integrations over a set of $u$ and $v$ variables on top of momentum (spin) integrations (sums), which is not a problem for Monte Carlo methods. However, the dependence of the integrand on Matsubara frequencies is again a {\it product of simple poles}, meaning that exact summation over all internal Matsubara indexes can be performed analytically and the result rotated to the real-frequency axis. [Writing all propagators and effective interactions in terms of spectral representations brings additional technical advantages, see \cite{SM}.]

Equations (\ref{Gspectral}), (\ref{Uspectral}) were used in Ref.~\cite{Holm98} for solving
the self-consistent GW-approximation at $T=0$. More importantly, spectral representation for the Green's function was employed in Ref.~\cite{ferreroT2} in the context of Anderson impurity model to compute the real-frequency response using analytic Matsubara integration. However, it was not realized that taken together Eqs.~(\ref{Cauchy})-(\ref{Uspectral}) offer a generic solution for obtaining real-frequency response in an arbitrary field-theoretical formulation of the interacting many body problem, including cases with frequency-dependent effective interactions.

The rest of this work is devoted to the explicit demonstration of how the proposed scheme works
in practice by considering the problem of the plasmon life-time in jellium.


{\it Starting point.} First, we need to construct $\tilde{G}$ and $U$. The jellium model is defined as the homogenous electron gas on a positive neutralizing background
\begin{equation}
H=\sum_i \frac{k_i^2}{2m} + \sum_{i<j} \frac{e^2}{|\mathbf{r}_i -\mathbf{r}_j|} - \mu N ,
\label{jellium}
\end{equation}
with $m$ the electron mass. In Fourier representation the bare interaction potential is given by
$V_0=4\pi e^2/q^2$. We use the inverse Fermi momentum, $1/k_F$, and Fermi energy, $\varepsilon_F = k_F^2/2m$, as units of length and energy, respectively, and employ the short-hand notation, $\sum_{\mathbf{k}} = (2\pi )^{-3} \int d^3k$, for momentum integrals. The definition of the Coulomb parameter $r_s$ in terms of the system number density, $\rho$, and Bohr radius is standard: $4\pi r_s^3/3 = 1/ \rho a_B^3$.

\begin{figure}[tbh]
\centerline{\includegraphics[angle = 0,width=0.95\columnwidth]{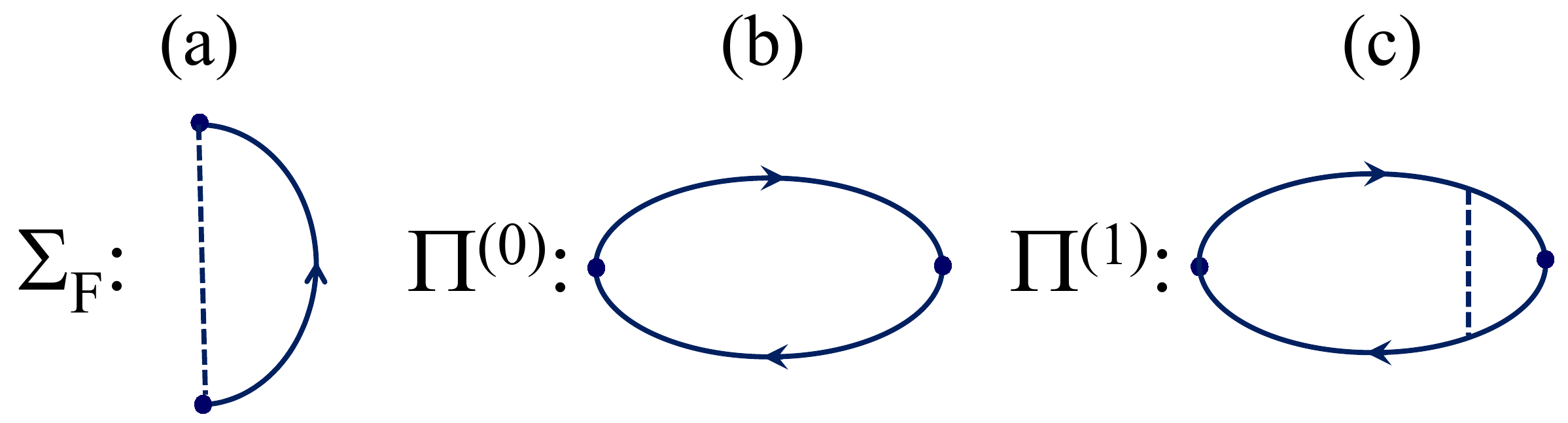}}
\caption{(color online) (a) Fock self-energy diagram; (b)-(c) $0$-th and $1$-st order (with respect to the number of interaction lines) contributions to the polarization operator $\Pi$.
Solid and dashed lines represent the fermionic propagators, $\tilde{G}$, and Yukawa potentials, $Y$, respectively.}
\label{Fig1}
\end{figure}

Following Ref.~\cite{Kun2019} in the Matsubara domain, we expand on top of the self-consistent Hartree-Fock solution for the Green's function, and tune the chemical potential to obtain the desired value of $r_s$. To avoid divergent Fermi-velocity renormalization, this solution is based on the
Yukawa potential, $Y(q)=4\pi e^2 /(q^2+\kappa^2)$, with appropriately chosen parameter $\kappa$ [see discussion below Eq.~(\ref{defF})]. To be specific, $\tilde{G}^{-1} (\mathbf{k}, i\omega_n) = i\omega_n - \epsilon_\mathbf{k}$, where the renormalized dispersion relation $\epsilon_\mathbf{k}= k^2/2m - \mu +\Sigma_{F}(\mathbf{k})$ is iterated using relations
$\Sigma_{F}(\mathbf{k}) = \sum_{\mathbf q} Y(\mathbf{q}) n(\epsilon_{\mathbf{k}-\mathbf{q}})$,
see Fig.~\ref{Fig1}(a), and $2 \sum_{\mathbf{k}} n(\epsilon_{\mathbf{k}}) =\rho$, until convergence.

The same Yukawa potential was used in Ref.~\cite{Kun2019} in place of the effective interaction potential in the full diagrammatic expansion. This choice is not suitable for our purposes because the plasmon is a collective excitation; an expansion in powers of $Y$ will fail to describe a basic process, $ P \to P + e-h $, where the plasmon is losing its energy by emitting electron-hole pairs, unless certain geometrical series are summed up to infinity.

\begin{figure}[tbh]
\subfigure{\includegraphics[scale=0.24]{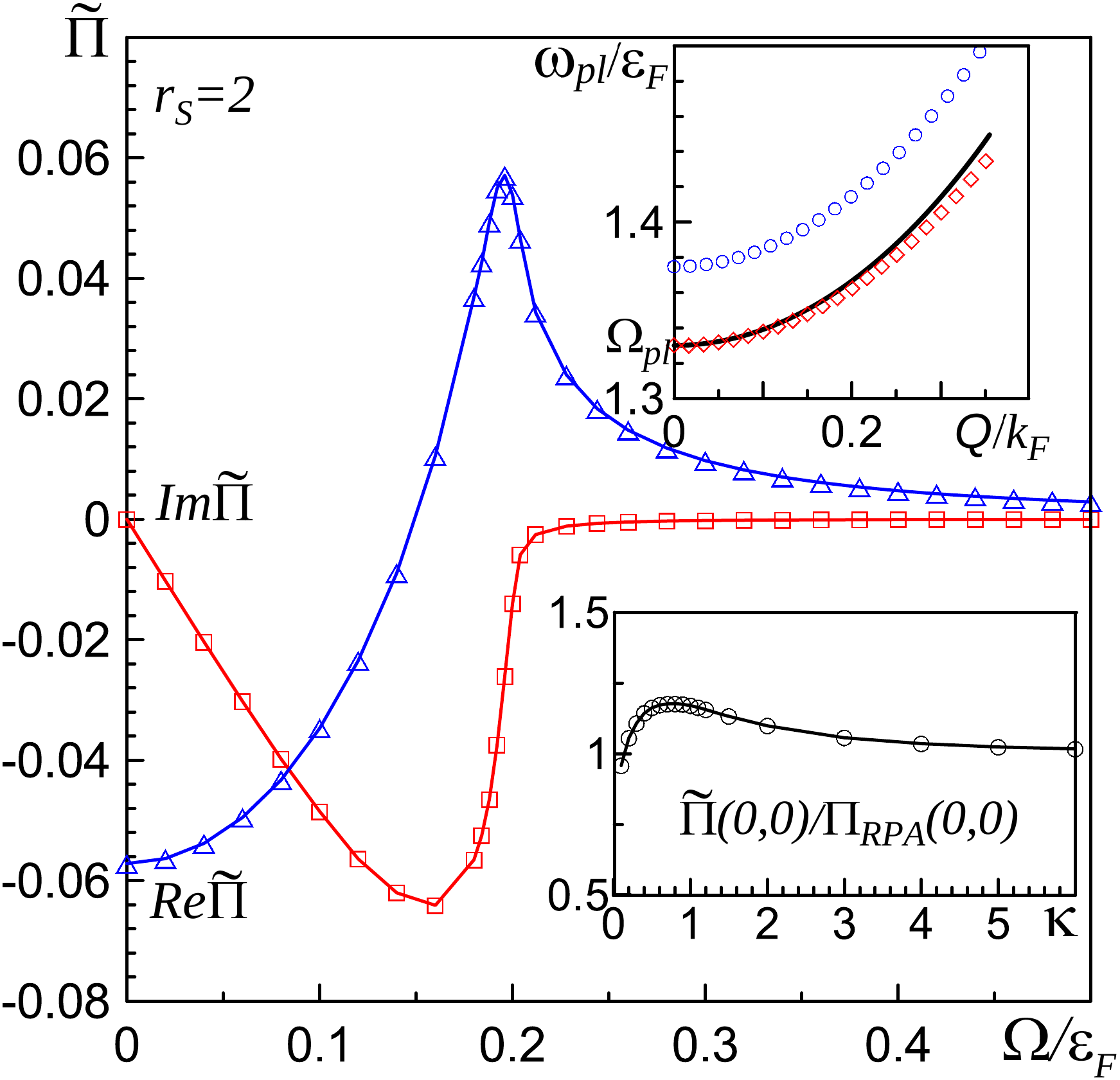}}
\subfigure{\includegraphics[scale=0.24]{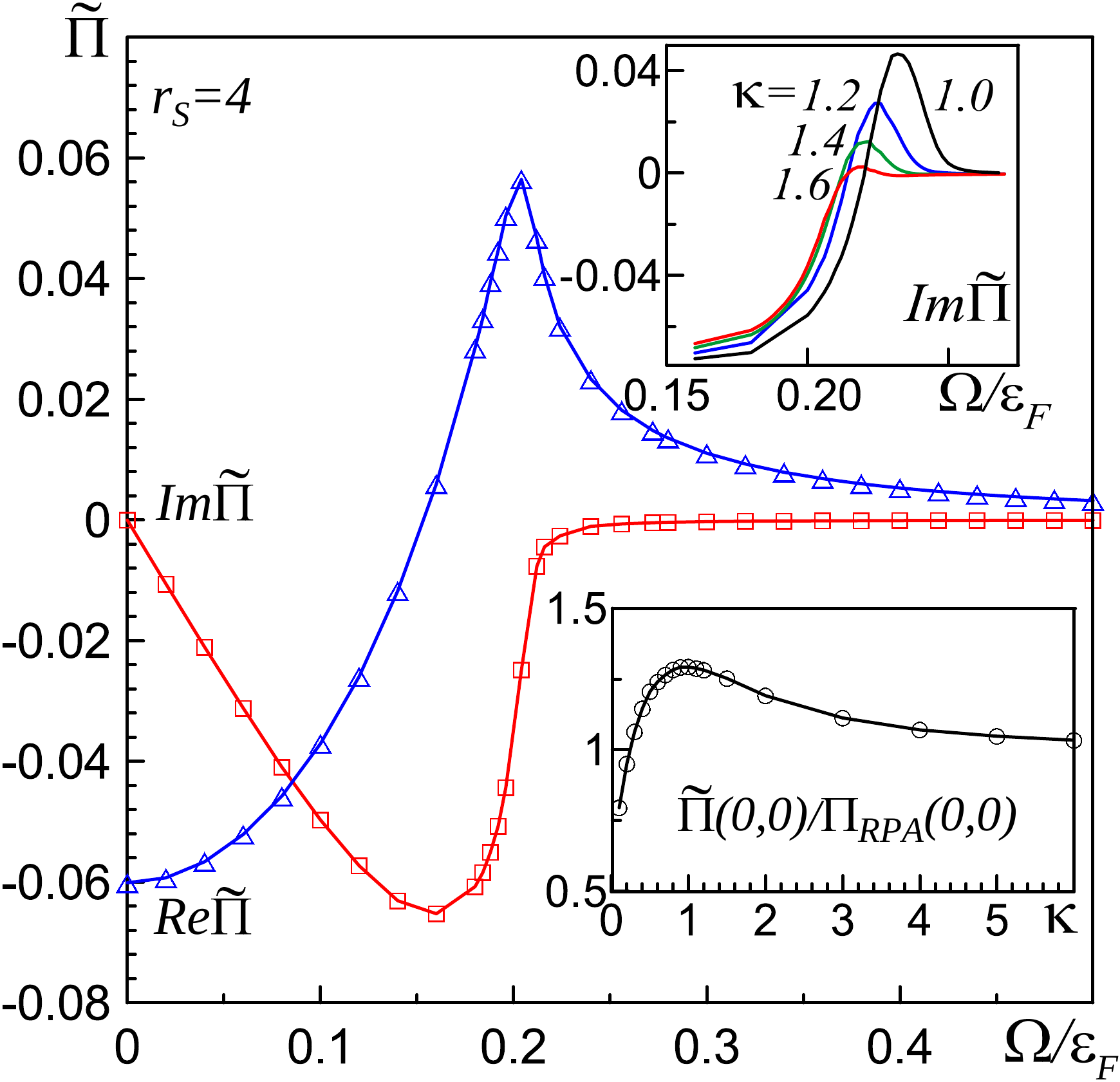}}
\caption{(color online) Polarization function, $\tilde{\Pi}=\Pi^{(0)}+\Pi^{(1)}$, dependence on frequency at low temperature $T/\epsilon_F=0.02$.
Left panel:  $r_s=2$, $Q/k_F=0.098437$, $\kappa/k_F=1.2$.
Right panel: $r_s=4$, $Q/k_F=0.103711$, $\kappa/k_F=1.8$.
Blue and red curves represent the real and imaginary parts of $\tilde{\Pi}$, respectively.
Lower insets show $\tilde{\Pi}(Q=0,\Omega=0)/\Pi_{RPA}(Q=0,\Omega=0)$ as a function of $\kappa$
for $r_s=2$ (left) and $r_s = 4$ (right). In the limit $\kappa \rightarrow \infty$ $\Pi(0, 0)$ saturates at $3 \rho / 2 \varepsilon_F$ equal to 0.05066 in our units.
The upper left inset shows the low-momentum part of the plasmon dispersion for $r_s=2$
within the (i) RPA (solid black curve),
           (ii)  $\tilde{\Pi}=\Pi^{(0)}$ approximation (blue circles), and
           (iii) $\tilde{\Pi}$ with vertex correction (red diamonds).
The upper right inset shows how $\rm{Im} \tilde{\Pi}$ for $r_s=4$ changes sign for $\kappa/k_F<1.8$.
All error bars are smaller than symbol sizes. }
\label{Fig2}
\end{figure}

These considerations force one to consider effective interactions based on dynamic screening effects
with ``built in" plasmon excitations (see also Section III of the Supplemental material \cite{SM}). It is tempting to start with $U^{-1}=V_0^{-1} - \Pi^{(0)}$, where $\Pi^{(0)}$ is given by the diagram shown in Fig.~\ref{Fig1}(b). However, the resulting plasmon spectrum (derived from the standard condition $U^{-1}(Q,\omega_{pl})=0$) strongly violates an exact hydrodynamic relation $ \omega_{pl}^2(Q=0)= \Omega_{pl}^2=4 \pi e^2 \rho /m$ for translation invariant systems. This problem is eliminated by adding the leading vertex correction, $U^{-1}=V_0^{-1} - \Pi^{(0)}-\Pi^{(1)}$, see Fig.~\ref{Fig1}(c). The upper inset in the left panel of Fig.~\ref{Fig2} shows that now the plasmon spectrum exhibits the proper behavior at $Q\to 0$, following the standard RPA result in this limit.

To complete the setup, we need to compute the spectral density $D$ in Eq.~(\ref{Uspectral}). For this it is sufficient to know the real and imaginary parts of $\tilde{\Pi}=\Pi^{(0)}+\Pi^{(1)}=R+iI$ on the
real frequency axis. If we split the total spectral density into the electron-hole continuum, $D_{e-h}$,
and the singular plasmon pole contribution, $D_{pl}$, then (see also Ref.~\cite{RIXS2020})
\bea
D_{e-h} &=& -\frac{I}{[(V^{-1}_0 - R)^2 + I^2]}; \label{DEH} \\
D_{pl}  &=& \pi r_{pl}(Q) \delta(\Omega - \omega_{pl}(Q)) \qquad \mbox{if} \qquad I(Q) = 0, \;\;\;
\label{DPL}
\eea
where $r_{pl} = 1/|\partial R /\partial\Omega |_{\omega_{pl}(Q)}$ is the pole residue.
After summation over Matsubara frequencies, the real-frequency result for $\tilde{\Pi}$ reads:
\be
\tilde{\Pi}= - 2\sum_{\bf p} {\cal F}_{{\bf p} + {\bf Q},{\bf p}} -
2\sum_{{\bf p},{\bf k}} Y({\bf p}-{\bf k}) {\cal F}_{{\bf p} + {\bf Q},{\bf p}} {\cal F}_{{\bf k}+{\bf Q},{\bf k}},
\label{Pi01}
\ee
\be
{\cal F}_{{\bf q}_1,{\bf q}_2} =
\frac{n_{\mathbf{q}_1}-n_{\mathbf{q}_2}} {\Omega - \epsilon_{\mathbf{q}_1} + \epsilon_{\mathbf{q}_2} + i \eta}.
\label{defF}
\ee

We evaluated momentum integrals in Eq.~(\ref{Pi01}) by standard Monte Carlo methods
on a dense mesh of $Q$ and $\Omega$ points for several values of $\kappa$. The optimized perturbation theory strategy \cite{Stevenson81,Feynman86} would be to choose $\kappa$ in such a way that the answer
computed up to a given order of expansion is least sensitive to its arbitrary value. Previous
work in this vein \cite{Kun2019} considered static properties only. For the fully dynamic calculation, one is further restricted by the condition that the spectral functions need to be positive for any $\Omega >0$. With respect to the low-frequency behavior, optimal values of $\kappa$ would correspond to the extrema of the $\tilde{\Pi}(0,0,\kappa)$ curves, shown in lower insets of Fig.~\ref{Fig2} for $r_s=2$ and $r_s=4$. The fact that both maxima are broad can be used to choose larger values of $\kappa$ without loss of accuracy in order to guarantee that $\rm{Im} \tilde{\Pi}(\Omega >0) <0$. Indeed, unless $\kappa$ is large enough, $\rm{Im} \tilde{\Pi}$ becomes positive in a finite frequency range, see the upper right inset in Fig.~\ref{Fig2}. Our strategy then is to choose large enough $\kappa$ as close as possible to the extremum of $\tilde{\Pi}(0,0,\kappa)$, leading to $\kappa/k_F = 1.2$ and $\kappa/k_F = 1.8$ for $r_s=2$ and $r_s=4$, respectively. The corresponding real and imaginary parts of $\tilde{\Pi}$ are presented in Fig.~\ref{Fig2}. At moderate values of $r_s$  the qualitative behavior remains similar to that in the RPA. The smoothing of the singularities in the polarization operator is a temperature effect.

\begin{figure}[tbh]
\centerline{\includegraphics[angle = 0,width=0.95\columnwidth]{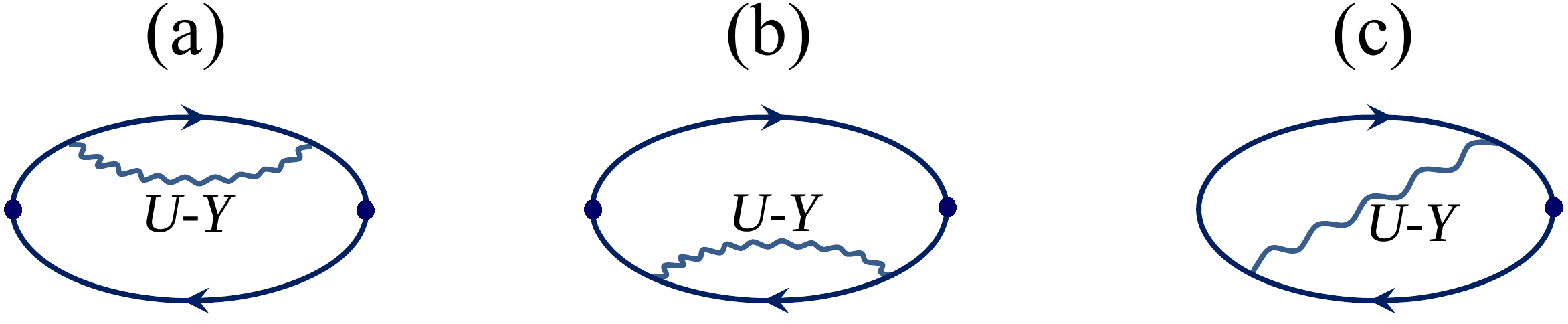}}
\caption{(color online) Lowest-order polarization diagrams within the formulation based on
$\tilde{G}$ and $U$.}
\label{Fig3}
\end{figure}

{\it Plasmon line-width.} In our formulation, the lowest-order polarization diagrams contributing to the finite plasmon life-time are shown in Fig.~\ref{Fig3}. To avoid double-counting, one has to subtract
Yukawa potentials from effective screened interactions, because the corresponding contributions
are already included in the definitions of $\tilde{G}$ and $U$ functions. The sum of all diagrams
in Fig.~\ref{Fig3} will be denoted as $\Delta \Pi$.

\begin{figure}[tbh]
\subfigure{\includegraphics[scale=0.24]{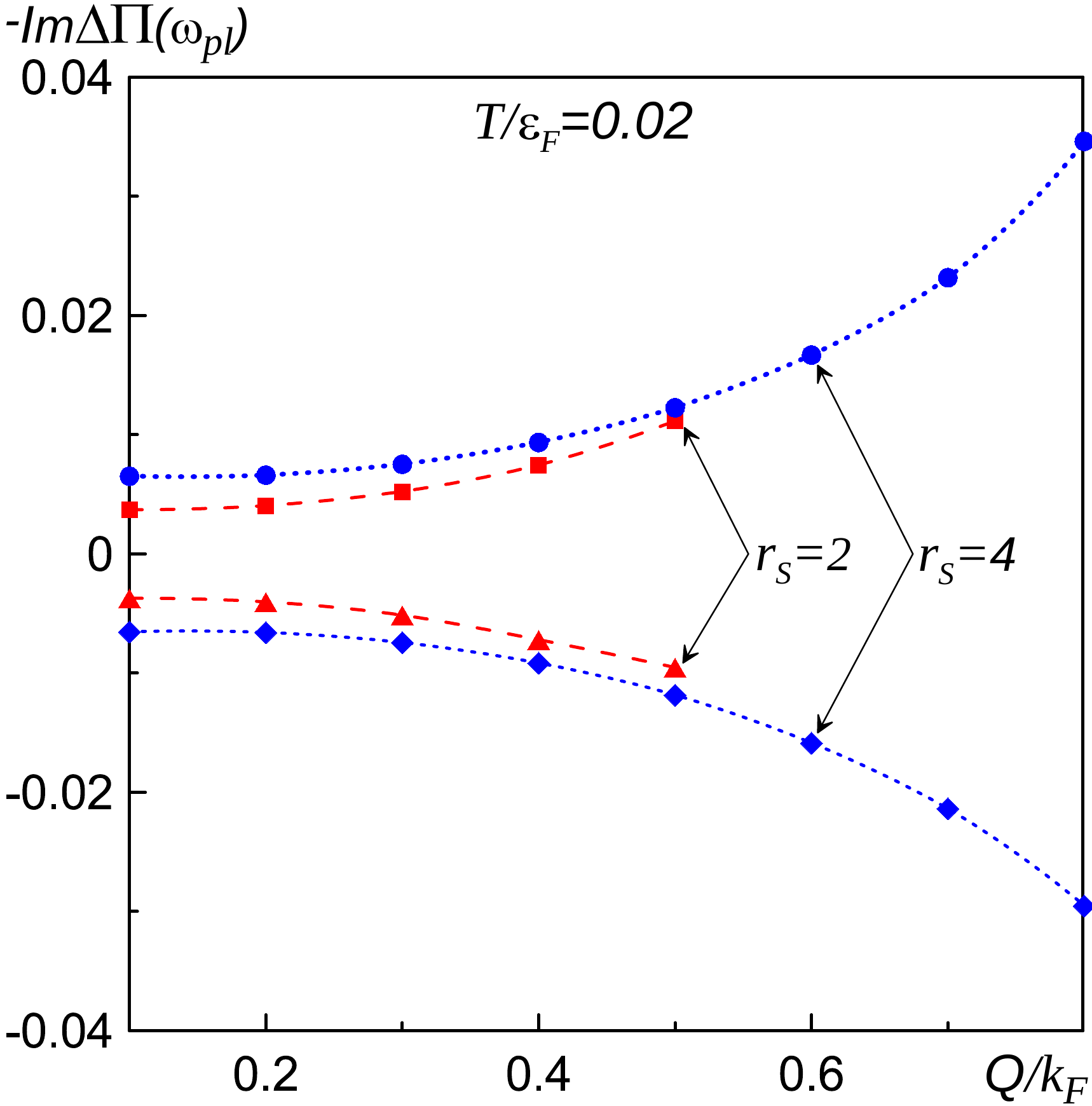}}
\subfigure{\includegraphics[scale=0.24]{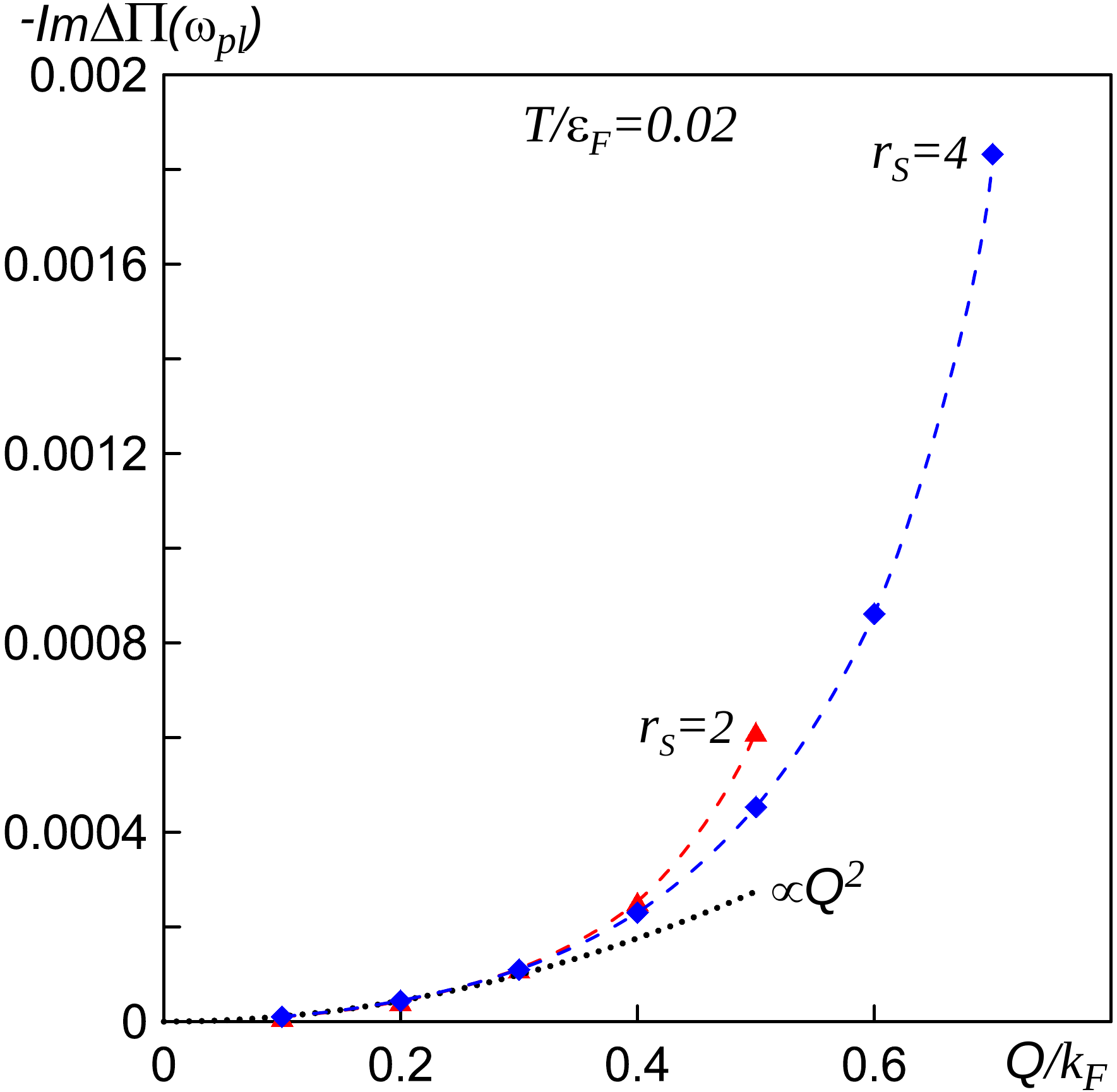}}
\caption{(color online) Minus imaginary part of the polarization operator contributions pictured
in Fig.~\ref{Fig3} as functions of momentum for $\Omega=\omega_{pl}(Q)$ and $T/\varepsilon_F=0.02$.
Left panel:  Upper curves are contributions from the sum of diagrams (a) and (b)
for $r_s=2$ (dashed with squares), and $r_s=4$ (dotted with circles).
             Lower curves are contributions from the diagram (c)
for $r_s=2$ (dashed with triangles), and $r_s=4$ (dotted with diamonds)
Right panel: The sum of diagrams (a), (b), and (c)
for $r_s=2$ (red dashed curve with triangles) and $r_s=4$ (blue dashed curve with diamonds).
The $Q^2$-dependence (black dotted curve) is added for comparison.
All error bars are smaller than symbol sizes. }
\label{Fig4}
\end{figure}

Accounting for the first two diagrams in Fig.~\ref{Fig3} would be equivalent to using the so-called
GW-approximation perturbatively. It is not surprising then that these two contributions strongly
violate another exact hydrodynamic condition, $\Pi (Q\to 0, \Omega \ne 0) \propto Q^2$, see Ref.~\cite{Pines}, because a similar situation takes place in the GW-approximation \cite{Holm98,JelGW}.
If we were to compute the plasmon line-width on the basis of the first two diagrams in Fig.~\ref{Fig3},
we would find that the plasmon excitation is completely destroyed at small momenta. Indeed, the data presented in the left panel of Fig.~\ref{Fig4} extrapolate to finite values at $Q=0$, leading to a divergent contribution after multiplication by the Coulomb potential (see also Fig.$4$ of the Supplemental material \cite{SM}).

It is thus crucial not to miss the vertex correction given by the diagram (c) in Fig.~\ref{Fig3}. It
compensates diagrams (a) and (b) almost perfectly for all values of $Q$, and restores the proper
$\propto Q^2 $ behavior of the (a)+(b)+(c) sum at small momenta, see right panel of Fig.~\ref{Fig4}. The involved analytical expressions for all diagrams (before Monte Carlo integration over internal momenta) can be found in the Supplemental material \cite{SM} (see Section II). While their derivation on the basis of Cauchy formula (\ref{Cauchy}) is straightforward, the number of terms rapidly increases with the number of frequency dependent lines, not to mention that $U$ functions (\ref{Uspectral}) contain three distinct contributions: frequency independent part, plasmon pole, and electron-hole continuum.

After evaluating the imaginary part of $\Delta \Pi (Q, \Omega=\omega_{pl}(Q))$ we
obtain the plasmon line-width from
\be
\gamma_{pl}(Q,T) = -r_{pl}(Q,T) \rm{Im} \Delta \Pi(Q,\omega_{pl}(Q,T),T).
\label{PLW}
\ee
[At small momenta $r_{pl} \approx V_0 \omega_{pl}/2 \propto Q^{-2}$.] Since the final result for $\gamma_{pl}$ is much smaller than $\omega_{pl}$ there is no need for performing a frequency scan. We have verified that the answer does not change when $Im \Delta \Pi$ is computed at frequencies $\omega_{pl} \pm \gamma_{pl}$.

\begin{figure}[tbh]
\subfigure{\includegraphics[scale=0.24]{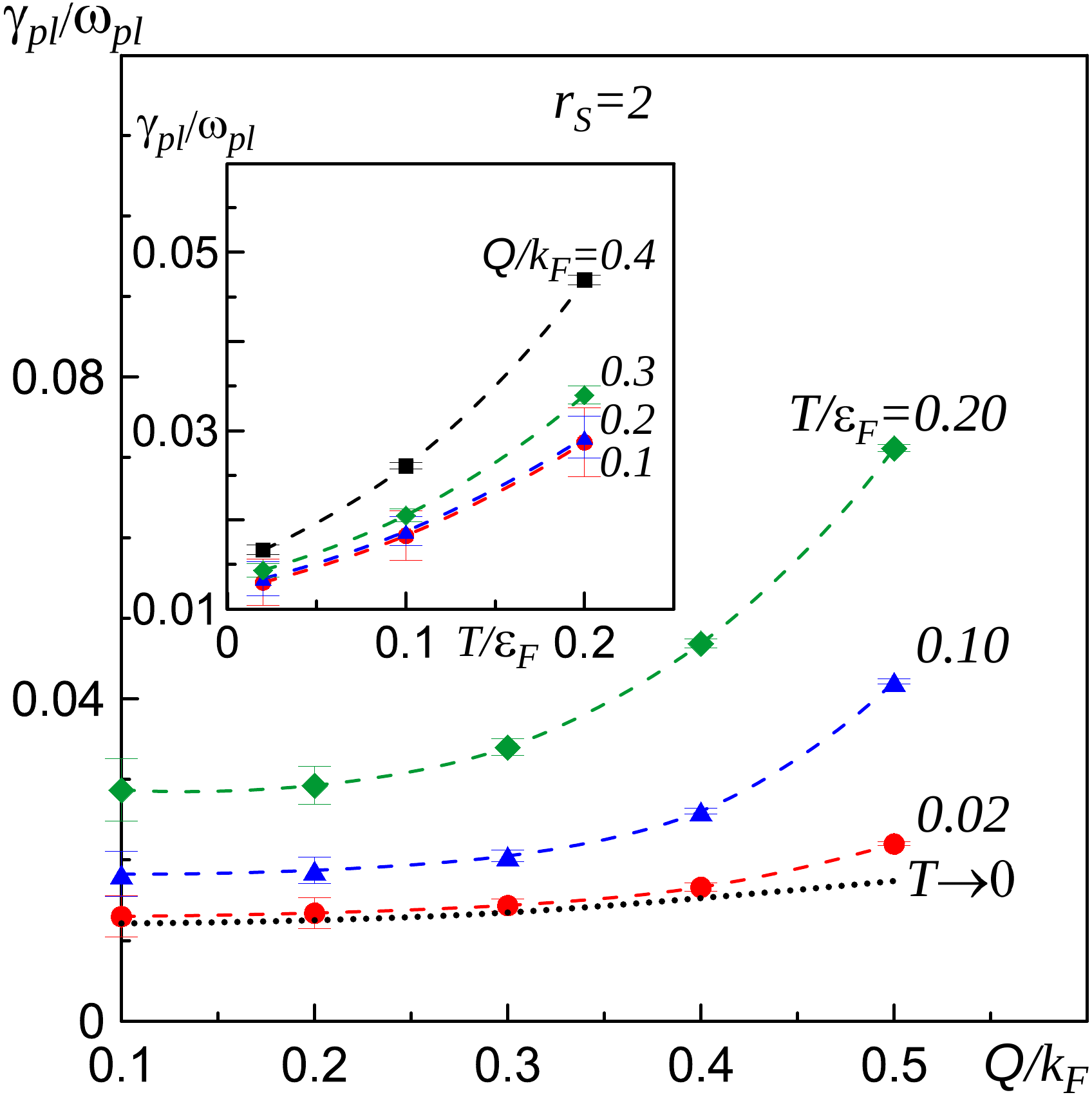}}
\subfigure{\includegraphics[scale=0.24]{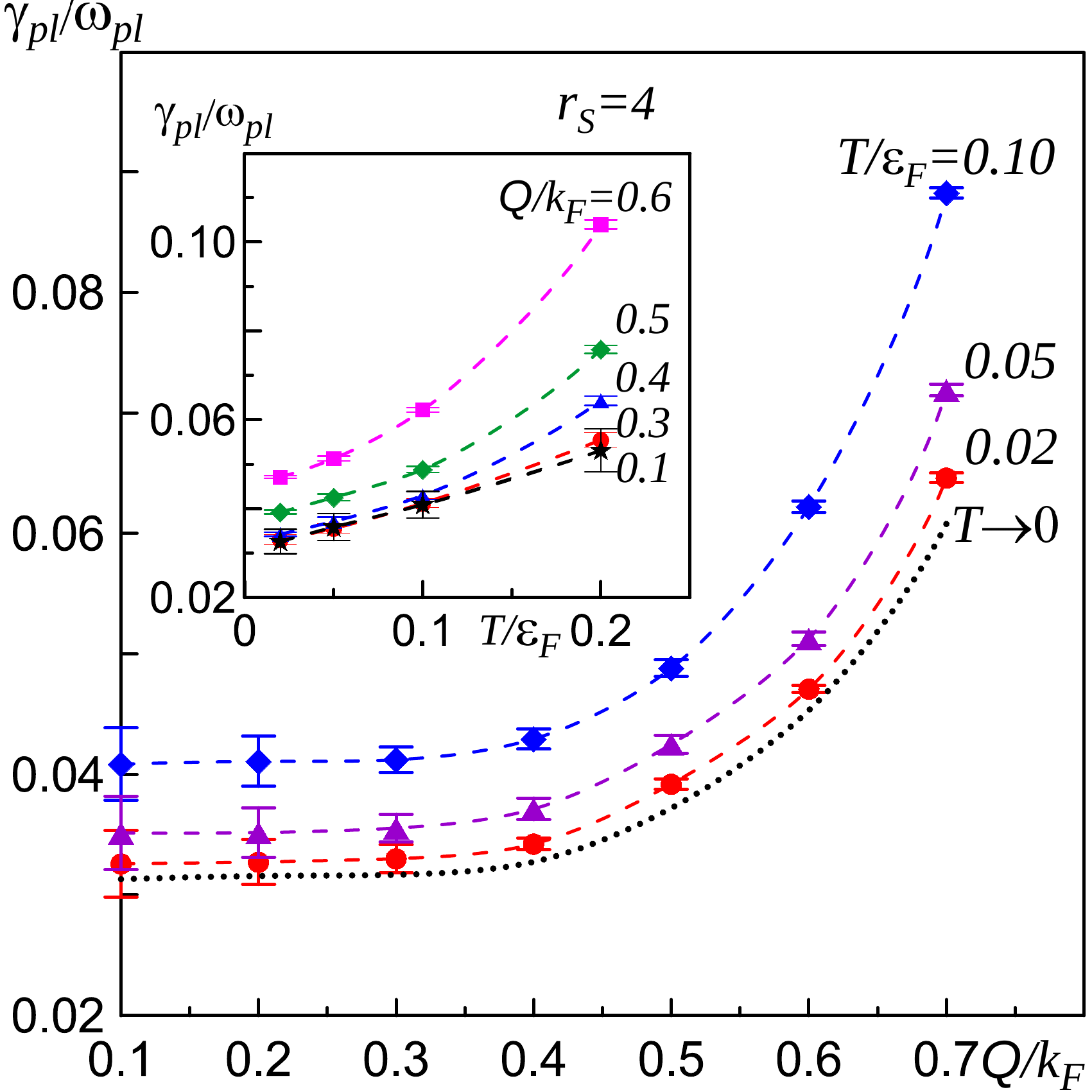}}
\caption{(color online) Plasmon line-width to plasmon frequency ratio as a function of momentum
at different temperatures for $r_s=2$ (left panel) and $4$ (right panel).
The black dotted lines in both panels show results extrapolated to the $T \to 0$ limit
using parabolic fits. Temperature dependence for different values of $Q$ is shown
in insets.}
\label{Fig5}
\end{figure}

Our final results for the plasmon line-width on the basis of diagrams with one $U$-line are discussed in Fig.~\ref{Fig5}. All data are presented as dimensionless ratios $\gamma_{pl}(Q,T)/\omega_{pl}(Q,T)$
to immediately see when plasmon excitations remain well-defined. This appears to be the case all the way to the plasmon spectrum end point for both values of $r_s$ when the temperature is low. The line-width saturates to a finite value in the $Q\to 0$ limit because the $Q^2$-dependence of $\rm{Im} \Delta \Pi$ is compensated by the divergence of the Coulomb potential present in the definition of the plasmon residue.

The answer is also finite in the $T\to 0$ limit. This can be understood on the basis of spectral density for two $(e-h)$ excitations that overlaps with the plasmon peak \cite{RIXS2020}. Thus there exist kinematically allowed decay channels for $Q=0$ plasmons excited from the ground state of the system. Somewhat surprising is the fact that the line-width remains rather small even for large vales of $r_s$. Finite-temperature corrections are linear at values $T/\varepsilon_F \ll 1$ with a much stronger temperature dependence emerging at $T/\varepsilon_F > 1$.

{\it Higher order contributions.} Our reformulation of the diagrammatic expansion in terms of $\tilde{G}$ and $U$ is exact, and one can proceed with computing higher-order diagrams using standard rules. We illustrate some of the second-order diagrams and process them in the Supplemental material \cite{SM} (see Section IV). While summation over internal Matsubara frequencies allows to perform calculations directly on the real-frequency axis, it also brings additional computational challenges. Using two next-order diagrams as an example, in \cite{SM} we demonstrate that processing Matsubara sums ``by hand" quickly leads to expressions of overwhelming complexity (the remaining momentum integrals are done by standard Monte Carlo techniques). Since the Cauchy formula (\ref{Cauchy}) is recursive, it should be possible to fully automate the process, similarly to what was done in Ref.~\cite{LeBlanc2019} for the case when only fermionic propagators were frequency dependent.

Given that certain groups of diagrams feature strong compensation, an efficient algorithm would need
to combine them analytically (see also Ref.~\cite{Kun2019}). We see clear advantages in implementing
the recursive scheme for obtaining Taylor expansions from skeleton diagrams \cite{semi-bold}, because
it automatically groups irreducible diagrams and reduces the number propagators and interaction lines.
This scheme also significantly simplifies processing of counter terms, and eliminates higher order poles in Mutsubara sums.

{\it Conclusion.}
Building on previous developments \cite{LeBlanc2019,Holm98,ferreroT2}, we
report a solution to the problem of computing finite-temperature response functions on the real frequency axis using Feynman diagrams for an arbitrary field-theoretical formulation of the interacting problem.  This includes problems with frequency-dependent effective interactions and dressed, renormalized, or self-consistent treatments required for producing convergent expansions. Spectral densities (of arbitrary complexity) for experimentally relevant observables (optical conductivity, resonant inelastic X-ray spectroscopy, neutron scattering, excitation life-times, etc.) can be computed with an accuracy that was never possible before.  Realistically, contribution from diagrams up to sixth order may be reached.

To illustrate how the technique works, we used it to compute the leading processes contributing to the finite plasmon line-with within the jellium model, and studied the line-width dependence on momentum and temperature for moderate values of the Coulomb parameter $r_s$. The increase of the interaction strength leads to a decrease of the plasmon life-time, but nevertheless the plasmon remains well defined.  One important qualitative result is the necessity to include vertex corrections in order to ensure the obtained results do not violate general principles. Future work will aim at developing efficient schemes for generating and processing real-frequency expressions for high-order diagrams to gain full control over systematic errors resulting from the series truncation.

{\it Acknowledgements.} A.M.T., R.M.K., and I.S.T. thank support from the Office of Basic Energy Sciences, Material Sciences and Engineering Division, U.S. Department of Energy under Contract No. DE-SC0012704. N.V.P. thanks support from the Simons Collaboration on the Many Electron Problem. The authors thank James LeBlank and Kun Chen for sharing details on their methods and helpful discussions.

\end{document}


\author{I. S. Tupitsyn}
\author{A. M. Tsvelik}
\author{R. M. Konik}
\author{N. V. Prokof'ev}


\title{Supplemental material for "Real-frequency response functions at finite temperature"}

\maketitle


\section{Analytic $\eta \to 0$ limit}
\label{ETA0}

One advantage of using spectral representations is that all poles are simple because
$\{ u,v \}$ variables are not subject to conservation laws, Ref.~\onlinecite{ferreroT2}.
The other advantage is that locations of all poles are given by simple
linear combinations of these variables. This is precisely what allows one to
perform the $\eta \to 0$ limit analytically, and make the entire scheme
exact. Let $s_i$ be one such linear combination and the corresponding pole
contribution to the answer has the form $ \int ds_i F(\dots ,s_i)/(\Omega-s_i+i\eta)$
(we suppress all other variables in $F$ for brevity). By taking the $\eta \to 0$ limit
using $1/(x+i\eta) = {\cal P}/x-i\pi \delta(x)$, were ${\cal P}$
stands for the principal value, we convert this expression into
\be
\int ds_i \bigg[ \frac{F(s_i)-F(2\Omega - s_i)}{2 (\Omega - s_i)}
- i \pi F(\Omega ) g(s_i) \bigg] .
\label{eta}
\ee
Here $g(s_i)$ is an arbitrary function normalized to unity. It eliminates the
need for tracking the remaining integration variables.

\section{Analysis of leading order diagrams in Fig. 3 of main text}
\label{UD}

\begin{figure}[tbh]
\centerline{\includegraphics[scale=0.4]{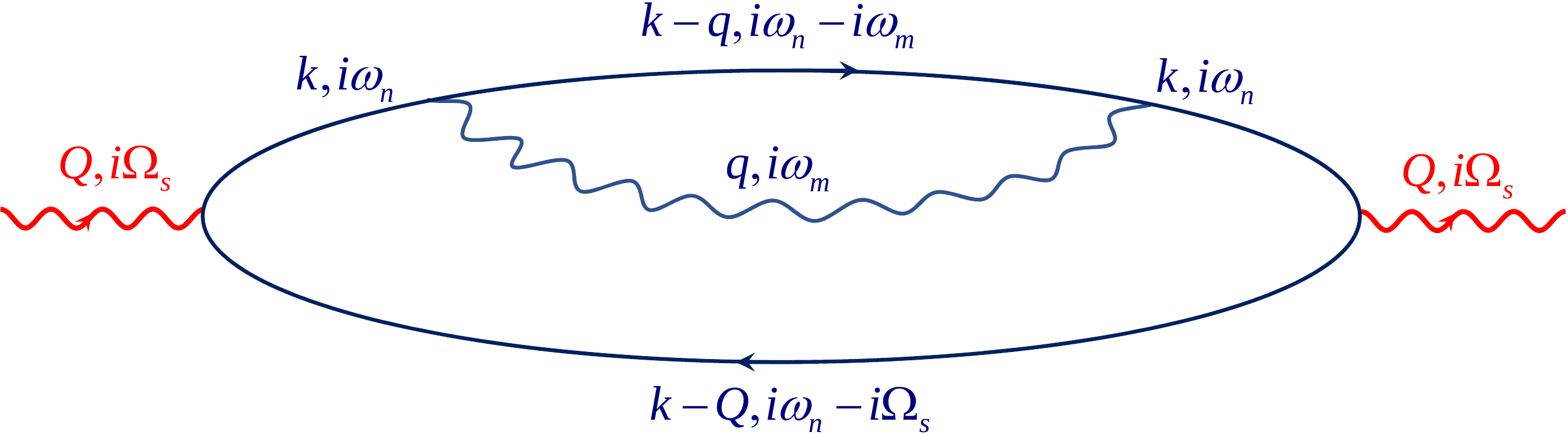}}
\caption{(color online) The same polarization diagram as in Fig.3(a) of the main text with all
frequency and momentum variables specified. $Q$ and $\Omega$ are considered to be external
parameters. Spin indices are not shown for brevity.}
\label{Fig1}
\end{figure}

Let us start with specifying all relevant external and internal variables for the diagram shown in Fig.3(a) of the main text, see Fig.~\ref{Fig1}.

\noindent
Before we proceed, let us introduce a convenient set of short-hand notations:
\[
\epsilon_{\mathbf{k},\mathbf{p}} = \epsilon(\mathbf{k}) - \epsilon(\mathbf{p})\,, \qquad
n_{u}=n(u)=\frac{1}{e^{u/T}+1}\,, \qquad
\delta n_{\mathbf{k},\mathbf{p}} = n_{\epsilon(\mathbf{k})} - n_{\epsilon(\mathbf{p})}\,, \qquad
n'_k = \frac{dn(x)}{dx} \vert_{x=\epsilon(\mathbf{k})} \,.
\]
\[
N_{u}=N(u)=\frac{1}{e^{u/T}-1}\,, \qquad
N_{\mathbf{k},\mathbf{p}} = N(\epsilon_{\mathbf{k},\mathbf{p}}) \,.
\]
Since the interaction line is given by
\[
U-Y = V_0(q)-Y(q) + (1/\pi) \int_{-\infty}^{\infty} du \frac{D(\mathbf{q},u)}{i\omega_m - u},
\]
we have two terms that need to be processed differently--the second
term has an extra pole. We thus define their contributions to $\Delta
\Pi^{(a)}$ via $\Delta \Pi^{(a)}=\Delta \Pi^{(a)}_{VY}+\Delta \Pi^{(a)}_{D}$.
A recursive application of the Cauchy formula for the $V_C(q)-Y(q)$ term results in
\be
\Delta \Pi^{(a)}_{VY} = -2 \sum_{{\bf k},{\bf q}} \frac{4 \pi e^2 \kappa^2}{q^2(q^2+\kappa^2)}
\frac{n_{\mathbf{k}-\mathbf{q}}} {-i\Omega_s-\epsilon_{\mathbf{k}-\mathbf{Q},\mathbf{k}}}
\bigg[ n'_k - \frac{\delta n_{\mathbf{k},\mathbf{k}-\mathbf{Q}}} {-i\Omega_s-\epsilon_{\mathbf{k}-\mathbf{Q},\mathbf{k}}} \bigg].
\label{UY-1}
\ee
For the second term we get:
\bea
\Delta \Pi^{(a)}_{D} &=& \frac{-2}{\pi} \sum_{{\bf k},{\bf q}} \int_{-\infty}^{\infty} du D(\mathbf{q},u)
\frac{(n_{\mathbf{k}-\mathbf{q}} - 1 - N_u)} {(i\Omega_s+\epsilon_{\mathbf{k}-\mathbf{Q},\mathbf{k}}) (\epsilon_{\mathbf{k}-\mathbf{q},\mathbf{k}} + u)} \\
&\times& \bigg[
n'_k + \frac{\delta n_{\mathbf{k},\mathbf{k}-\mathbf{Q}}} {i\Omega_s+\epsilon_{\mathbf{k}-\mathbf{Q},\mathbf{k}}}
+ \frac{ n_{\mathbf{k}} - n(\epsilon_{\mathbf{k}-\mathbf{q}} + u) }
{\epsilon_{\mathbf{k}-\mathbf{q},\mathbf{k}} + u}
+ \frac{n_{\mathbf{k}-\mathbf{Q}} - n(\epsilon_{\mathbf{k}-\mathbf{q}} + u)} {i\Omega_s-\epsilon_{\mathbf{k}-\mathbf{q},\mathbf{k}-\mathbf{Q}} - u }
\bigg].
\label{GW1-2}
\eea
Wick rotation to the real-frequency domain is now performed by a simple substitution
$i\Omega_s \to \Omega + i\eta$. For diagram (b) in Fig.3 of the main text the analytical
expression for $\Delta \Pi^{(b)}$ is obtained from that of $\Delta \Pi^{(a)}$ by the replacement $\Omega + i\eta \to -\Omega -i\eta$. For diagram (c) the final result for $\Delta \Pi^{(c)}=\Delta \Pi^{(c)}_{VY}+\Delta \Pi^{(c)}_{D}$ reads
\be
\Delta \Pi^{(c)}_{VY} = -2 \sum_{{\bf k},{\bf q}}
\frac{4 \pi e^2 \kappa^2}{q^2(q^2+\kappa^2)}
\frac{\delta n_{\mathbf{k}, \mathbf{k}-\mathbf{Q}}} {-\Omega -i\eta -\epsilon_{\mathbf{k}-\mathbf{Q},\mathbf{k}}}
\frac{\delta n_{\mathbf{k}-\mathbf{q}, \mathbf{k}-\mathbf{q}-\mathbf{Q}}} {-\Omega -i\eta -\epsilon_{\mathbf{k}-\mathbf{q}-\mathbf{Q},\mathbf{k}-\mathbf{q}}},
\label{UY-3}
\ee
\bea
\Delta \Pi^{(c)}_{D} &=& \frac{-2}{\pi} \sum_{{\bf k},{\bf q}} \int_{-\infty}^{\infty} du
\frac{D(\mathbf{q},u)}{(-\Omega -i\eta-\epsilon_{\mathbf{k}-\mathbf{Q},\mathbf{k}})
(-\Omega -i\eta-\epsilon_{\mathbf{k}-\mathbf{q}-\mathbf{Q},\mathbf{k}-\mathbf{q}})}
\\
&\times& \bigg\{
(n_{\mathbf{k}-\mathbf{q}} - 1 - N_u) \bigg[
\frac{n_{\mathbf{k}-\mathbf{Q}}-n(\epsilon_{\mathbf{k}-\mathbf{q}}+u)}
{-\Omega -i\eta+\epsilon_{\mathbf{k}-\mathbf{q},\mathbf{k}-\mathbf{Q}}+u}
-\frac{n_{\mathbf{k}}-n(\epsilon_{\mathbf{k}-\mathbf{q}}+u)}
{\epsilon_{\mathbf{k}-\mathbf{q},\mathbf{k}}+u} \bigg] \nonumber \\
&& - (n_{\mathbf{k}-\mathbf{q}-\mathbf{Q}} - 1 - N_u) \bigg[
\frac{n_{\mathbf{k}-\mathbf{Q}}-n(\epsilon_{\mathbf{k}-\mathbf{q}-\mathbf{Q}}+u)}
{\epsilon_{\mathbf{k}-\mathbf{q}-\mathbf{Q},\mathbf{k}-\mathbf{Q}}+u}
-\frac{n_{\mathbf{k}}-n(\epsilon_{\mathbf{k}-\mathbf{q}-\mathbf{Q}}+u)}
{\Omega +i\eta+\epsilon_{\mathbf{k}-\mathbf{q}-\mathbf{Q},\mathbf{k}}+u} \bigg]
\bigg\}. \nonumber
\label{YV-4}
\eea

\section{Second order diagrams based on Yukawa potential and counter terms}
\label{Yukawa}

\begin{figure}[h]
\subfigure{\includegraphics[scale=0.425]{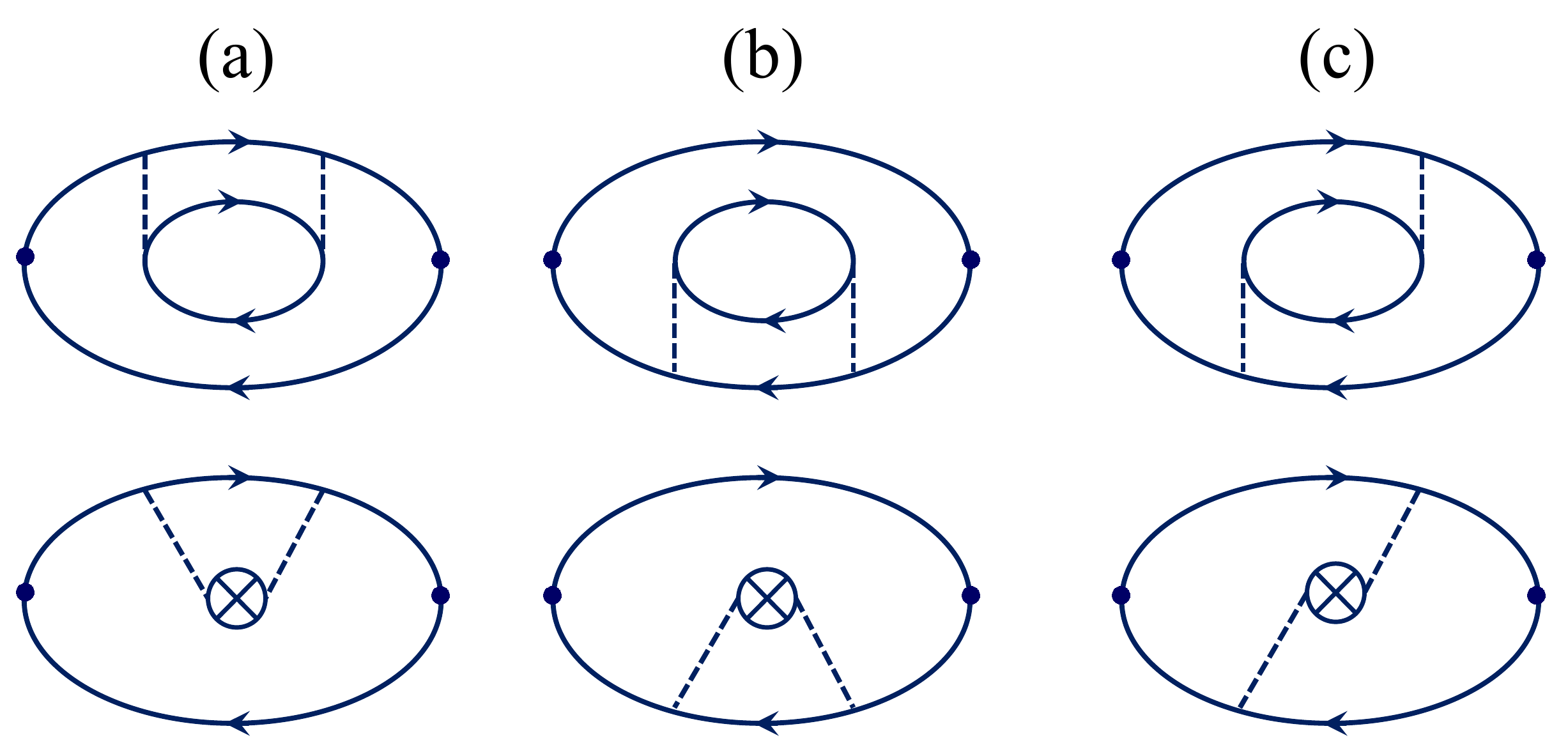}}
\caption{Upper row: Second-order (with respect to the number of interaction lines) diagrams
for the polarization operator with self-energy insertions on the upper (a) and lower (b)
propagator lines and vertex correction (c). The counter-terms are shown in the lower row.
Dashed lines represent the Yukawa potential.}
\label{Fig2}
\end{figure}

One may wonder whether having a ``built in" plasmon excitation and more elaborate field-theoretical formulation has a significant effect on the final result. We can answer in the affirmative here. To see this we consider effective interactions based on the same Yukawa potential as in the Hartree-Fock treatment of $\tilde{G}$ (this choice was made in Ref.~\onlinecite{Kun2019} which provides further details about the expansion) and compute contributions from diagrams having the same structure as diagrams in Fig.~3 of the main text. These diagrams are obtained by replacing $U-Y$ lines with the bubble diagram and the associated counter-term, see Fig.~\ref{Fig2}.

\begin{figure}[tbh]
\subfigure{\includegraphics[scale=0.325]{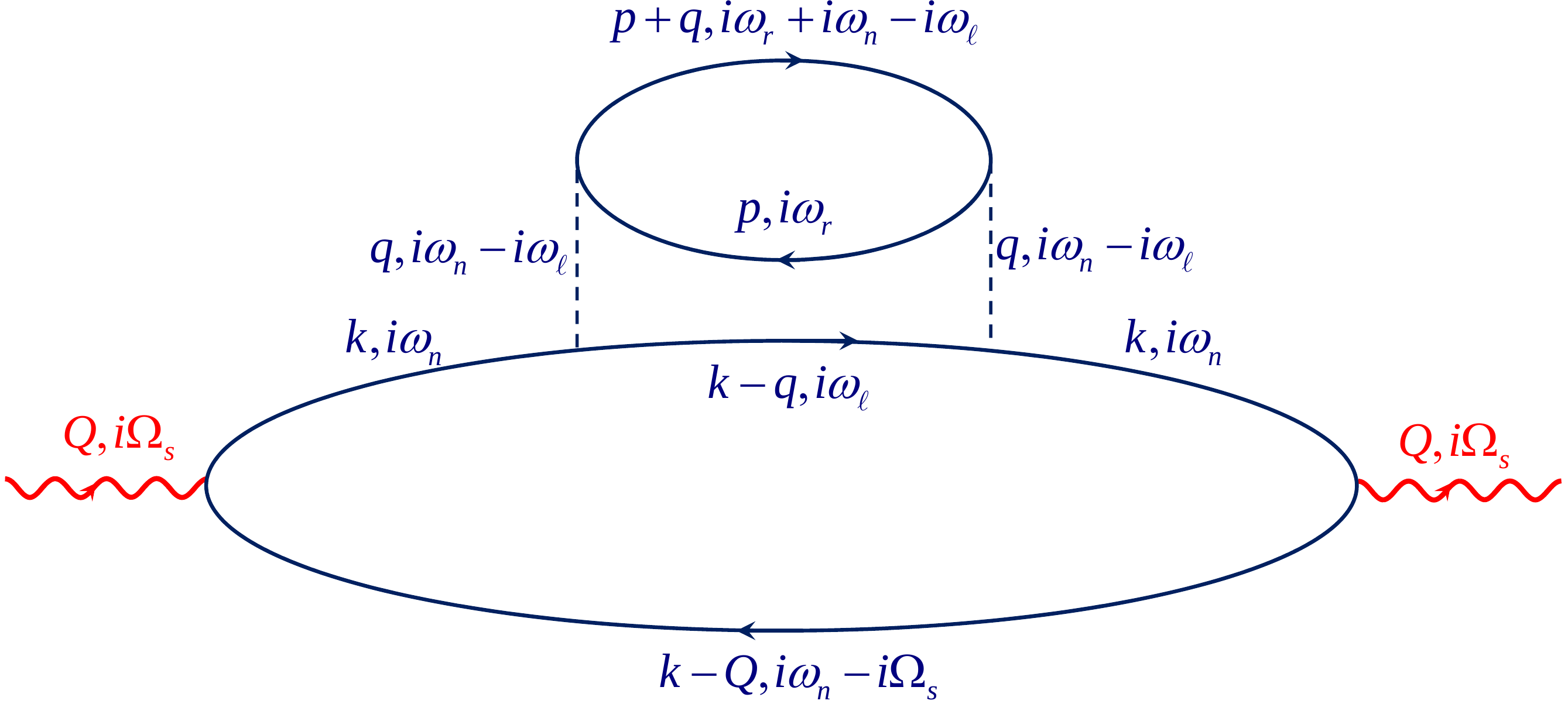}}
\subfigure{\includegraphics[scale=0.325]{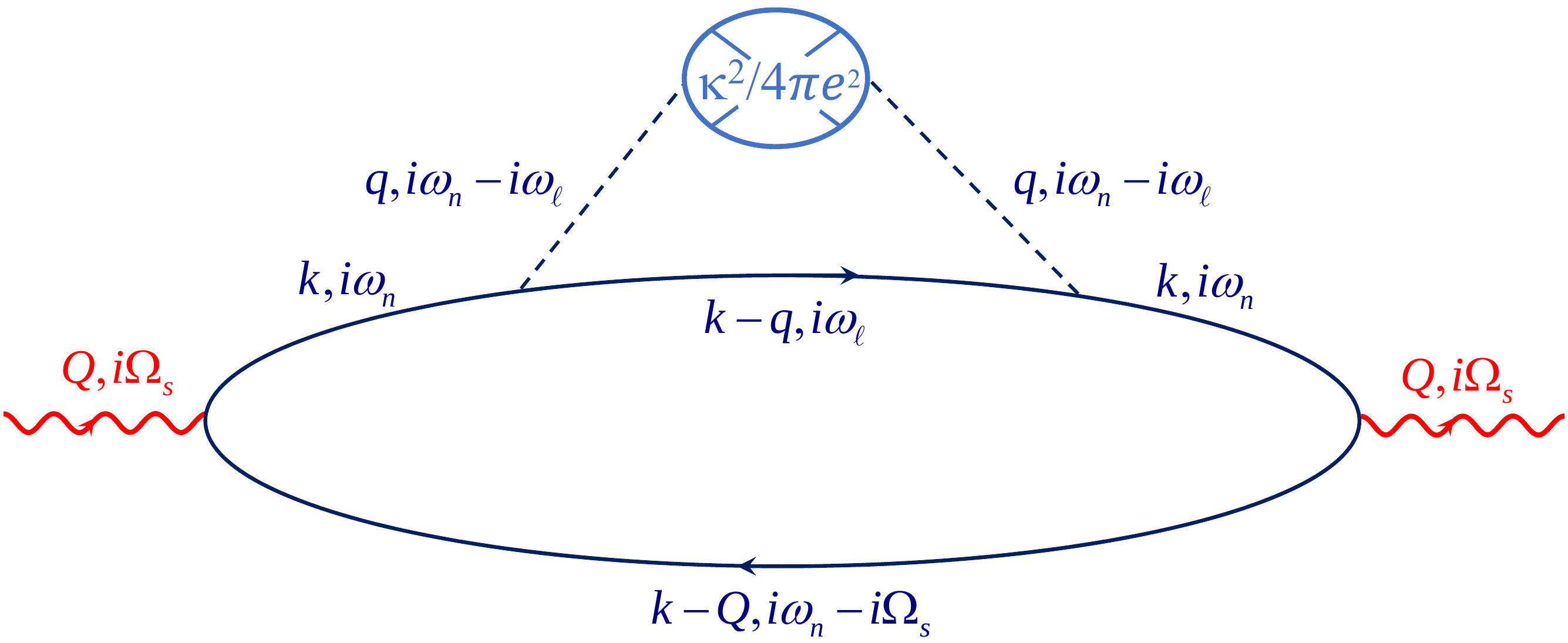}}
\caption{(color online) Second-order (with respect to the number of interaction lines) diagram
for polarization operator with self-energy insertion (left); the counter-term is shown to the right.
Dashed lines represent the Yukawa potential.}
\label{Fig3}
\end{figure}
Fig. \ref{Fig3} here provides all details for the diagrams in Fig.~\ref{Fig2}(a). The counter-term (CT) contribution is the easiest one:
\be
\Pi^{(a)}_{CT} = 2 \frac{\kappa^2}{4 \pi e^2}
\sum_{{\bf k}, {\bf q}} Y^2(\mathbf{q})
\frac{n_{\mathbf{k}-\mathbf{q}}} {-i\Omega_s-\epsilon_{\mathbf{k}-\mathbf{Q},\mathbf{k}}}
\bigg[ n'_k - \frac{\delta n_{\mathbf{k},\mathbf{k}-\mathbf{Q}}} {-i\Omega_s-\epsilon_{\mathbf{k}-\mathbf{Q},\mathbf{k}}} \bigg].
\label{GW1-1}
\ee
The main diagram has two additional poles and an extra momentum integration:
\bea
\Pi^{(a)}_{MD} &=& -4 \sum_{{\bf k},{\bf p},{\bf q}} Y^2(\mathbf{q})
\frac{(n_{\mathbf{k}-\mathbf{q}} + n_{\mathbf{p}+\mathbf{q}} -1) (n_{\mathbf{p}} + N(\epsilon_{\mathbf{p}+\mathbf{q}} +\epsilon_{\mathbf{k}-\mathbf{q}}))} {(-i\Omega_s-\epsilon_{\mathbf{k}-\mathbf{Q},\mathbf{k}}) (\epsilon_{\mathbf{p}+\mathbf{q},\mathbf{p}} + \epsilon_{\mathbf{k}-\mathbf{q},\mathbf{k}})} \\
&\times& \bigg[
\frac{ n(\epsilon_{\mathbf{p}+\mathbf{q}} + \epsilon_{\mathbf{k}-\mathbf{q}}) - n_{\mathbf{k}}}
{\epsilon_{\mathbf{p}+\mathbf{q},\mathbf{p}} + \epsilon_{\mathbf{k}-\mathbf{q},\mathbf{k}}}
- n'_k + \frac{\delta n_{\mathbf{k},\mathbf{k}-\mathbf{Q}}} {-i\Omega_s-\epsilon_{\mathbf{k}-\mathbf{Q},\mathbf{k}}}
- \frac{n(\epsilon_{\mathbf{p}+\mathbf{q},\mathbf{p}} + \epsilon_{\mathbf{k}-\mathbf{q}})
- n_{\mathbf{k}-\mathbf{Q}}  }{-i\Omega_s+\epsilon_{\mathbf{p}+\mathbf{q},\mathbf{p}} + \epsilon_{\mathbf{k}-\mathbf{q},\mathbf{k}-\mathbf{Q}}}
\bigg].
\label{GW1-2}
\eea
The complementary expression for diagrams in Fig.~\ref{Fig2}(b) is obtained from (\ref{GW1-2})
after Wick rotation by the replacement $\Omega + i\eta \to -\Omega -i\eta$.
Diagrams in Fig.~\ref{Fig2}(c) are computed following the same protocol.

\begin{figure}[tbh]
\subfigure{\includegraphics[scale=0.4]{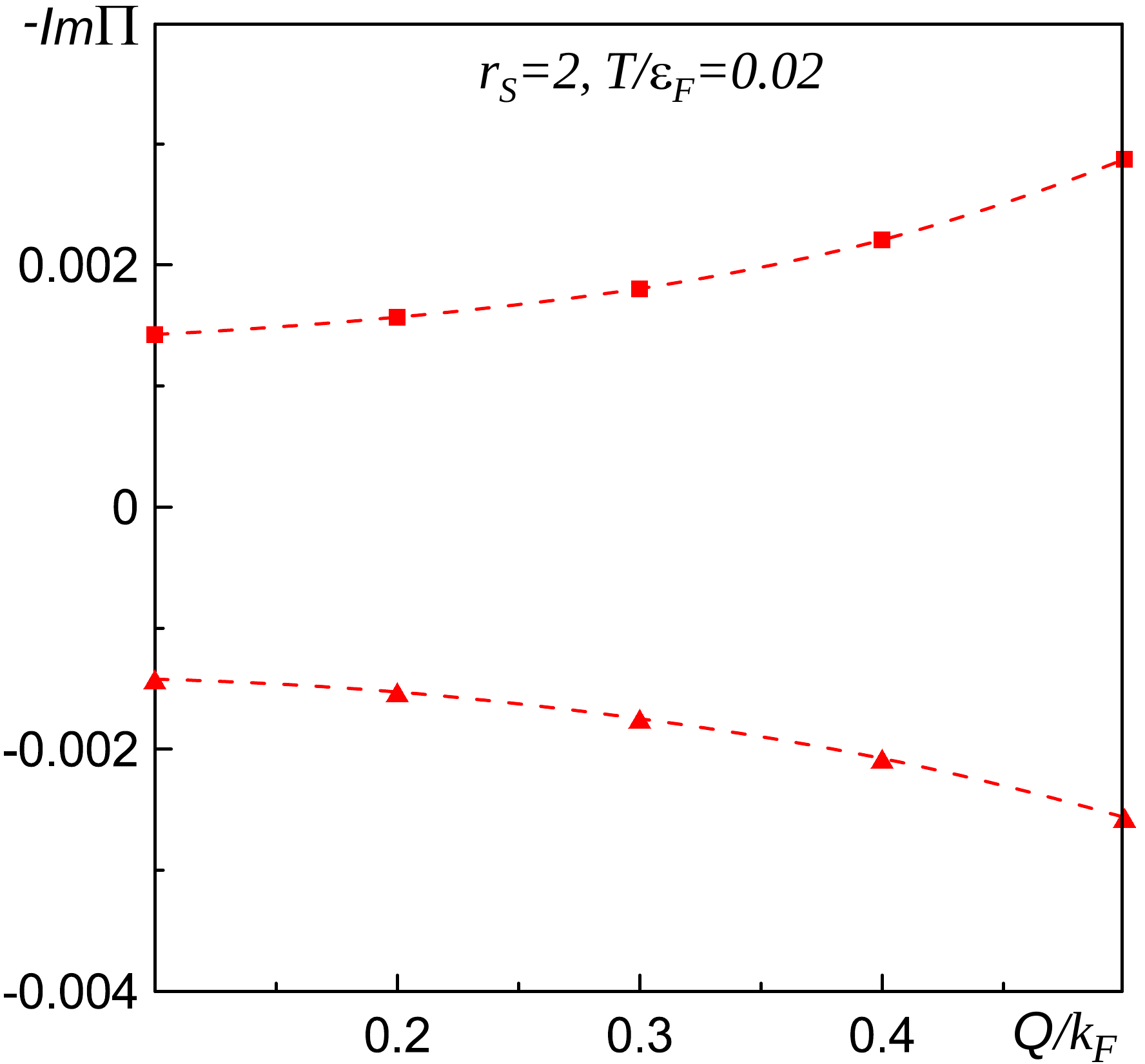}}
\subfigure{\includegraphics[scale=0.4]{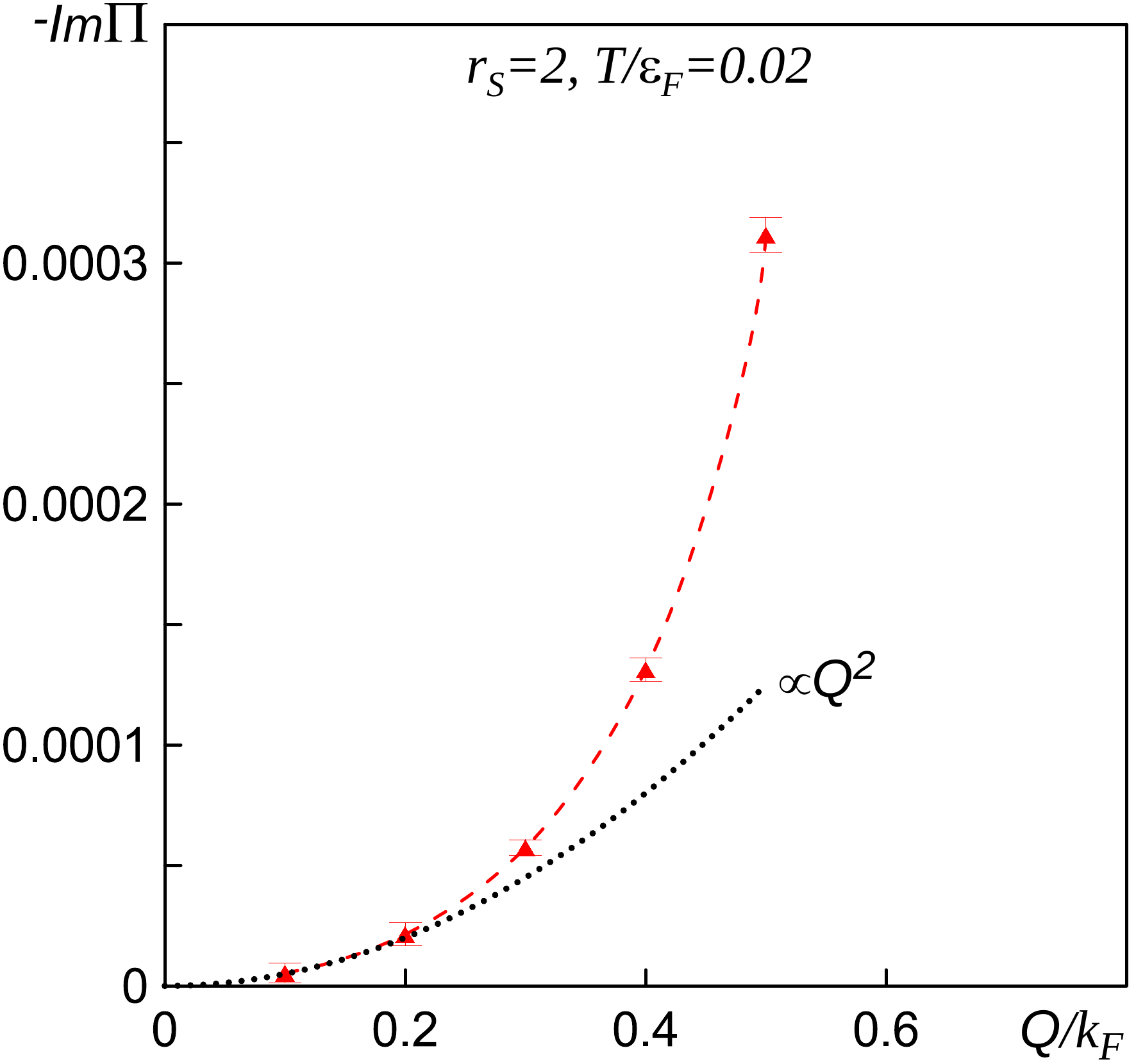}}
\caption{(color online) Left panel: Minus imaginary part of partial contributions to the polarization operator from diagrams in Fig.~\ref{Fig2} as a functions of momentum
at $r_s=2$, $T/\varepsilon_F=0.02$, and $\Omega=\omega_{pl}(Q)$.
The sum of diagrams (a) and (b) is shown by red squares. The vertex correction (c) is shown by red up-triangles. Right panel: The sum of all three diagrams (red up-triangles) is compared to the $Q^2$ fit (dotted black curve).}
\label{Fig4}
\end{figure}

In Fig.~\ref{Fig4} we present results based on calculations of diagrams (a), (b), and (c) in Fig.~\ref{Fig2} at $r_s=2$ and $T/\varepsilon_F=0.02$ for $\Omega=\omega_{pl}(Q)$.
Taken separately, these diagrams are physically meaningless because they saturate to a constant
value at $Q\to 0$, and have opposite sign. Only the sum of all three diagrams leads to the
proper $Q^2$ behavior at small $Q$. Quantitatively, these diagrams predict the plasmon line-width
that is nearly a factor of two smaller than results presented in the main text,
demonstrating the importance of constructing  an expansion with ``built in" collective excitations.

\section{Second-order diagrams with $U$ potentials}
\label{2ndO}

\begin{figure}[tbh]
\centerline{\includegraphics[scale=0.425]{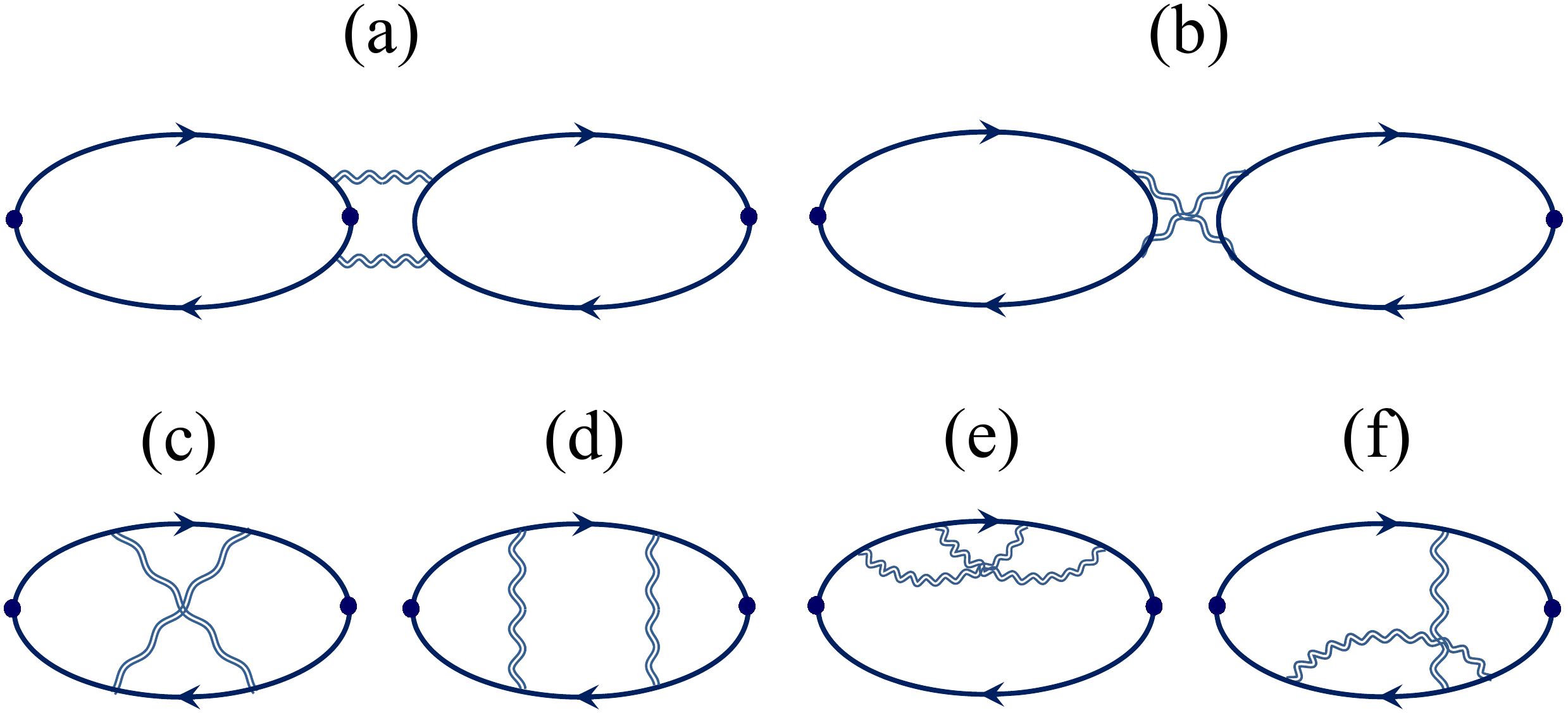}}
\caption{(color online) Examples of second order polarization diagrams in the new formulation.
To differentiate between the $U-Y$ functions that take care of double-counting and $U$, we
used single-wavy lines for the former in the main text and double-wavy lines for the latter here.}
\label{Fig5}
\end{figure}

In Fig.~\ref{Fig5} we list some of the second-order diagrams that appear in the expansion based on
$\tilde{G}$ and $U$ functions.
In what follows we demonstrate that higher-order real-frequency calculations
for expansions based on retarded effective interactions are feasible. In this
work we deal with all Matsubara sums by hand. The reader can immediately see that
the complexity of expressions quickly becomes overwhelming. With additional
algorithmic developments, this task can be fully automated.

\begin{figure}[tbh]
\subfigure{\includegraphics[scale=0.325]{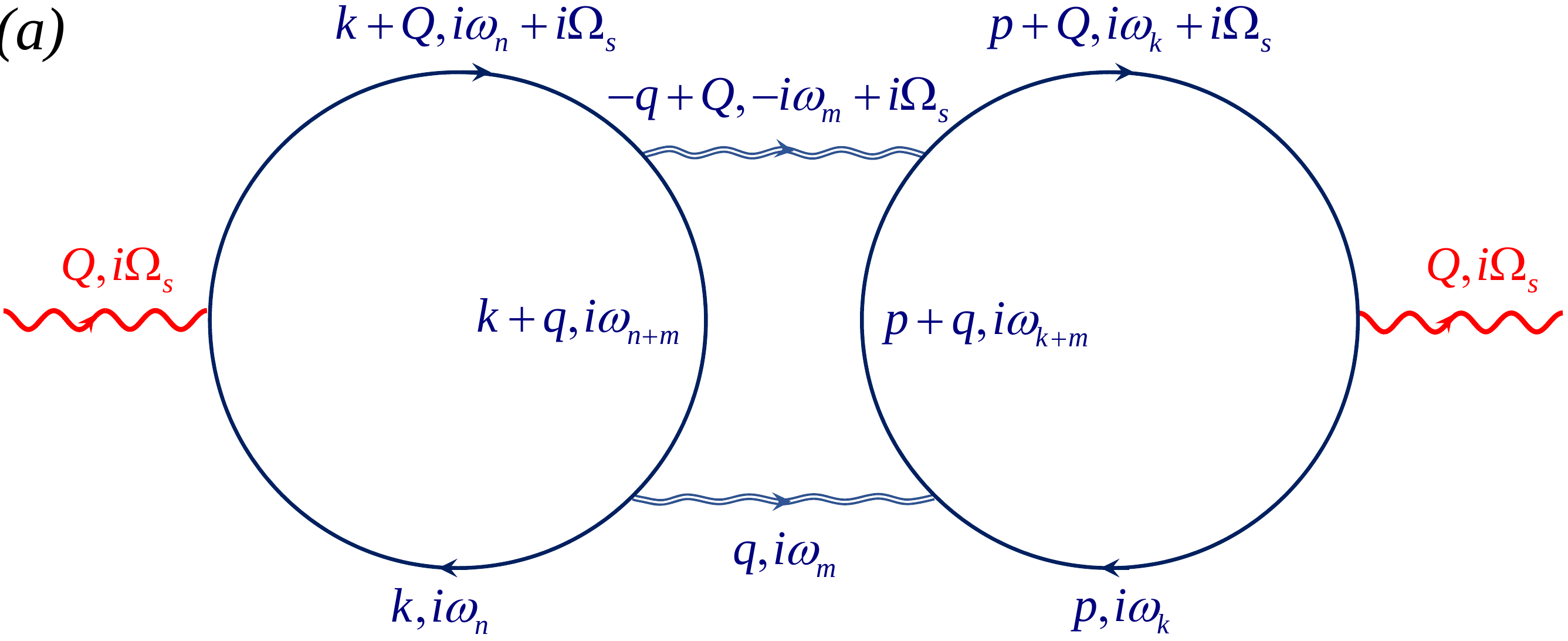}}
\subfigure{\includegraphics[scale=0.325]{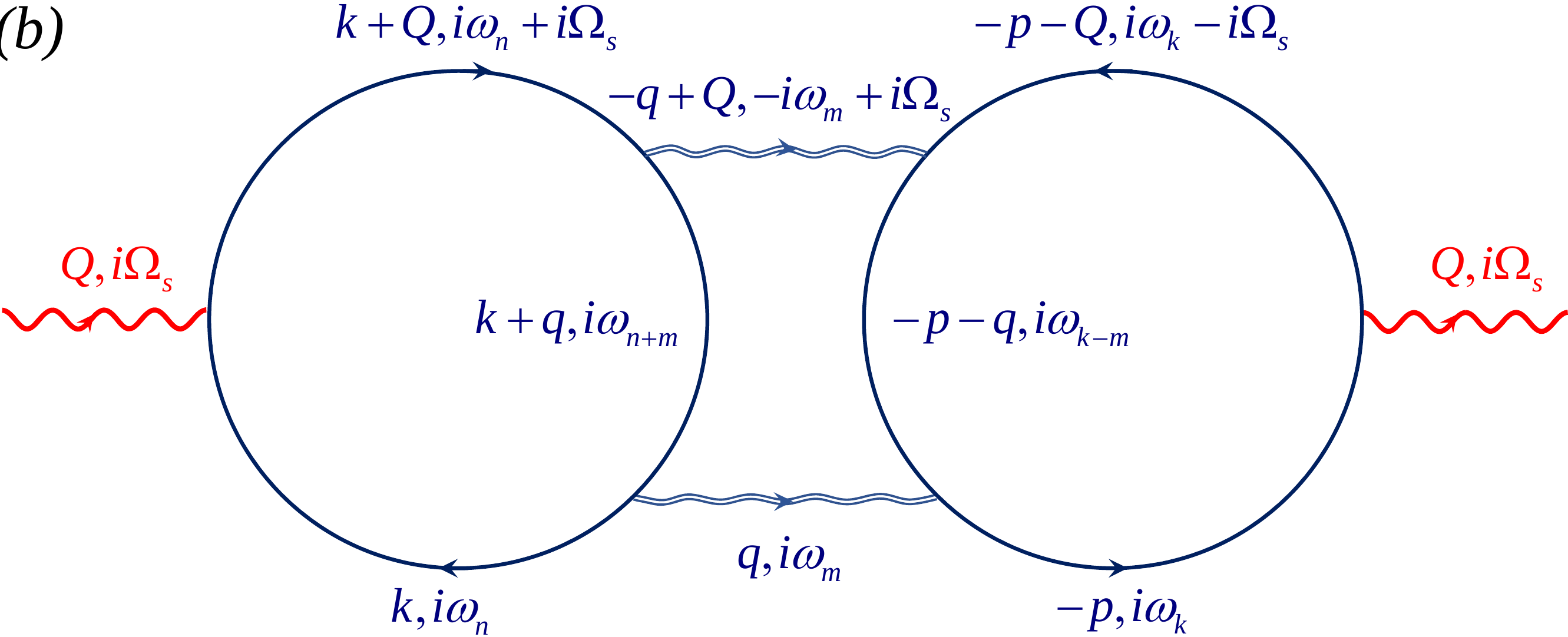}}
\caption{(color online) Second order polarization diagrams with two fermionic loops.
For convenience, in the diagram $(b)$ we flip the sign of the momentum variable for the right loop and
use reflection symmetry $\epsilon(-\mathbf{p}) = \epsilon(\mathbf{p}) = \epsilon(p)$.
Note that $\omega_m$ and $\Omega_s$ are bosonic Matsubara frequencies.}
\label{Fig6}
\end{figure}
As an example, we compute the first two diagrams shown in Fig.~\ref{Fig5}.
Their complete characterization is given in Fig.~\ref{Fig6}. First we performs sums
over indexes $n$ and $k$ to obtain contributions associated with the left and right loops.
For the left loops we have
\be
f_L(\mathbf{k}) =
\frac{1}{i\Omega_s-\epsilon_{\mathbf{k}+\mathbf{Q},\mathbf{k}}}
\left[ \frac{\delta n_{\mathbf{k},\mathbf{k}+\mathbf{q}}}
{i\omega_m-\epsilon_{\mathbf{k}+\mathbf{q},\mathbf{k}}} -
\frac{\delta n_{\mathbf{k}+\mathbf{Q},\mathbf{k}+\mathbf{q}}} {i\omega_m-i\Omega_s-\epsilon_{\mathbf{k}+\mathbf{q},\mathbf{k}+\mathbf{Q}}}
\right].
\label{DiagA-L}
\ee
The right loop for diagram (a) makes an identical contribution with $\mathbf{k} \to \mathbf{p}$.
For the right loop in diagram (b) we get
\be
f_R(\mathbf{p}) =
\frac{1}{-i\Omega_s-\epsilon_{\mathbf{p}+\mathbf{Q},\mathbf{p}}}
\left[ \frac{\delta n_{\mathbf{p},\mathbf{p}+\mathbf{q}}}
{-i\omega_m -\epsilon_{\mathbf{p}+\mathbf{q},\mathbf{p}}}
 - \frac{\delta n_{\mathbf{p}+\mathbf{Q},\mathbf{p}+\mathbf{q}}}
{-i\omega_m+i\Omega_s-\epsilon_{\mathbf{p}+\mathbf{q},\mathbf{p}+\mathbf{Q}}}
\right].
\ee
Given that each $U$ function has two dissimilar terms, there are four contributions to the polarization operator:
$\Pi^{(2)}_{a+b} = -4 \sum_{\mathbf{k}, \mathbf{p},\mathbf{q}} [C_{1}+C_{2}+C_{3}+C_{4}]$, where
\bea
C_{1} &=&  V_C(\mathbf{Q}-\mathbf{q}) V_C(\mathbf{q})
T \sum_m {\cal F};
\label{PiA} \\
C_{2} &=&  \frac{1}{\pi} \int_{-\infty}^{\infty} du V_C(\mathbf{Q}-\mathbf{q}) D(\mathbf{q},u)\,
T \sum_m \frac{{\cal F}}{i\omega_m -u};
\label{PiB} \\
C_{3} &=& \frac{1}{\pi} \int_{-\infty}^{\infty} du D(\mathbf{Q}-\mathbf{q},u) V_C(\mathbf{q}) \,
T \sum_m \frac{\cal F}{-i\omega_m + i\Omega_s -u} ;
\label{PiC} \\
C_{4} &=& \frac{1}{\pi^2} \int_{-\infty}^{\infty} du \int_{-\infty}^{\infty} dv
D(\mathbf{Q}-\mathbf{q},u) D(\mathbf{q},v) \,
T \sum_m \frac{\cal F}{(-i\omega_m + i\Omega_s -u)(i\omega_m - v)}  ; \;\;\;\;\;
\label{PiD} \\
{\cal F} &=& f_L(\mathbf{k},i\omega_m) [f_L(\mathbf{p},i\omega_m) + f_R(\mathbf{p},i\omega_m)].
\label{FLR}
\eea

Sums over $m$ lead to functions $S_1$, $S_2$, $S_3$, and $S_4$ described below.
They have to be substituted into Eqs.~(\ref{PiA}-\ref{PiD}) and the result
further integrated over internal momenta to obtain $\Pi^{(2)}_{a+b}$.

\underline{$S_{1}$}. For diagram (a), $S_1$ is given by the product of
$[ (i\Omega_s-\epsilon_{\mathbf{k}+\mathbf{Q},\mathbf{k}})
(i\Omega_s-\epsilon_{\mathbf{p}+\mathbf{Q},\mathbf{p}}) ]^{-1}$ and
\bea
&& \bigg\{ \frac{\delta n_{\mathbf{k},\mathbf{k}+\mathbf{q}} \delta n_{\mathbf{p},\mathbf{p}+\mathbf{q}}
[N_{\mathbf{p}+\mathbf{q},\mathbf{p}} - N_{\mathbf{k}+\mathbf{q},\mathbf{k}}]}
{\epsilon_{\mathbf{k}+\mathbf{q},\mathbf{k}} - \epsilon_{\mathbf{p}+\mathbf{q},\mathbf{p}}}
+ \frac{\delta n_{\mathbf{k}+\mathbf{Q},\mathbf{k}+\mathbf{q}} \delta n_{\mathbf{p}+\mathbf{Q},\mathbf{p}+\mathbf{q}}
[N_{\mathbf{p}+\mathbf{q},\mathbf{p}+\mathbf{Q}} - N_{\mathbf{k}+\mathbf{q},\mathbf{k}+\mathbf{Q}}]}
{\epsilon_{\mathbf{k}+\mathbf{q},\mathbf{k}+\mathbf{Q}} - \epsilon_{\mathbf{p}+\mathbf{q},\mathbf{p}+\mathbf{Q}}} - \nonumber \\
&&\frac{\delta n_{\mathbf{k},\mathbf{k}+\mathbf{q}} \delta n_{\mathbf{p}+\mathbf{Q},\mathbf{p}+\mathbf{q}}
[N_{\mathbf{p}+\mathbf{q},\mathbf{p}+\mathbf{Q}} - N_{\mathbf{k}+\mathbf{q},\mathbf{k}}]}
{-i\Omega_s+\epsilon_{\mathbf{k}+\mathbf{q},\mathbf{k}} - \epsilon_{\mathbf{p}+\mathbf{q},\mathbf{p}+\mathbf{Q}}}
- \frac{\delta n_{\mathbf{k}+\mathbf{Q},\mathbf{k}+\mathbf{q}} \delta n_{\mathbf{p},\mathbf{p}+\mathbf{q}}
[N_{\mathbf{p}+\mathbf{q},\mathbf{p}} - N_{\mathbf{k}+\mathbf{q},\mathbf{k}+\mathbf{Q}}]}
{i\Omega_s+\epsilon_{\mathbf{k}+\mathbf{q},\mathbf{k}+\mathbf{Q}} - \epsilon_{\mathbf{p}+\mathbf{q},\mathbf{p}}} \bigg\}.
\eea

Similarly, for diagram (b), $S_1$ is given by the product of
$[ (i\Omega_s-\epsilon_{\mathbf{k}+\mathbf{Q},\mathbf{k}})
(-i\Omega_s-\epsilon_{\mathbf{p}+\mathbf{Q},\mathbf{p}}) ]^{-1}$ and
\bea
&& \bigg\{ \frac{\delta n_{\mathbf{k},\mathbf{k}+\mathbf{q}} \delta n_{\mathbf{p},\mathbf{p}+\mathbf{q}}
[1+N_{\mathbf{p}+\mathbf{q},\mathbf{p}} + N_{\mathbf{k}+\mathbf{q},\mathbf{k}}]}
{\epsilon_{\mathbf{k}+\mathbf{q},\mathbf{k}} + \epsilon_{\mathbf{p}+\mathbf{q},\mathbf{p}}}
+ \frac{\delta n_{\mathbf{k}+\mathbf{Q},\mathbf{k}+\mathbf{q}} \delta n_{\mathbf{p}+\mathbf{Q},\mathbf{p}+\mathbf{q}}
[1+N_{\mathbf{p}+\mathbf{q},\mathbf{p}+\mathbf{Q}} + N_{\mathbf{k}+\mathbf{q},\mathbf{k}+\mathbf{Q}}]}
{\epsilon_{\mathbf{k}+\mathbf{q},\mathbf{k}+\mathbf{Q}} + \epsilon_{\mathbf{p}+\mathbf{q},\mathbf{p}+\mathbf{Q}}} - \nonumber \\
&&\frac{\delta n_{\mathbf{k},\mathbf{k}+\mathbf{q}} \delta n_{\mathbf{p}+\mathbf{Q},\mathbf{p}+\mathbf{q}}
[1+N_{\mathbf{p}+\mathbf{q},\mathbf{p}+\mathbf{Q}} + N_{\mathbf{k}+\mathbf{q},\mathbf{k}}]}
{-i\Omega_s+\epsilon_{\mathbf{k}+\mathbf{q},\mathbf{k}} + \epsilon_{\mathbf{p}+\mathbf{q},\mathbf{p}+\mathbf{Q}}}
- \frac{\delta n_{\mathbf{k}+\mathbf{Q},\mathbf{k}+\mathbf{q}} \delta n_{\mathbf{p},\mathbf{p}+\mathbf{q}}
[1+N_{\mathbf{p}+\mathbf{q},\mathbf{p}} + N_{\mathbf{k}+\mathbf{q},\mathbf{k}+\mathbf{Q}}]}
{i\Omega_s+\epsilon_{\mathbf{k}+\mathbf{q},\mathbf{k}+\mathbf{Q}} + \epsilon_{\mathbf{p}+\mathbf{q},\mathbf{p}}} \bigg\}.
\eea

\vspace{5mm}

\underline{$S_{2}$}. For diagram $(a)$, $S_2$ is given by the product of
$[ (i\Omega_s-\epsilon_{\mathbf{k}+\mathbf{Q},\mathbf{k}})
(i\Omega_s-\epsilon_{\mathbf{p}+\mathbf{Q},\mathbf{p}}) ]^{-1}$ and
\bea
\bigg\{
\frac{\delta n_{\mathbf{k},\mathbf{k}+\mathbf{q}} \delta n_{\mathbf{p},\mathbf{p}+\mathbf{q}}}
{\epsilon_{\mathbf{k}+\mathbf{q},\mathbf{k}} - \epsilon_{\mathbf{p}+\mathbf{q},\mathbf{p}}}
&\times&\left[
\frac{N_{\mathbf{k}+\mathbf{q},\mathbf{k}} - N_{u}}
        {u-\epsilon_{\mathbf{k}+\mathbf{q},\mathbf{k}}}
 - \frac{N_{\mathbf{p}+\mathbf{q},\mathbf{p}} - N_{u}}
        {u-\epsilon_{\mathbf{p}+\mathbf{q},\mathbf{p}}}
\right] + \nonumber \\
\frac{\delta n_{\mathbf{k}+\mathbf{Q},\mathbf{k}+\mathbf{q}}
\delta n_{\mathbf{p}+\mathbf{Q},\mathbf{p}+\mathbf{q}}}
{\epsilon_{\mathbf{k}+\mathbf{q},\mathbf{k}+\mathbf{Q}} - \epsilon_{\mathbf{p}+\mathbf{q},\mathbf{p}+\mathbf{Q}}}
&\times&\left[
   \frac{N_{\mathbf{k}+\mathbf{q},\mathbf{k}+\mathbf{Q}} - N_{u}}
   {-i\Omega_s+u-\epsilon_{\mathbf{k}+\mathbf{q},\mathbf{k}+\mathbf{Q}}}
 - \frac{N_{\mathbf{p}+\mathbf{q},\mathbf{p}+\mathbf{Q}} - N_{u}}
   {-i\Omega_s+u-\epsilon_{\mathbf{p}+\mathbf{q},\mathbf{p}+\mathbf{Q}}}
\right] - \nonumber \\
\frac{\delta n_{\mathbf{k},\mathbf{k}+\mathbf{q}}
\delta n_{\mathbf{p}+\mathbf{Q},\mathbf{p}+\mathbf{q}}}
{-i\Omega_s+\epsilon_{\mathbf{k}+\mathbf{q},\mathbf{k}} - \epsilon_{\mathbf{p}+\mathbf{q},\mathbf{p}+\mathbf{Q}}}
&\times&\left[
   \frac{N_{\mathbf{k}+\mathbf{q},\mathbf{k}} - N_{u}}
        {u-\epsilon_{\mathbf{k}+\mathbf{q},\mathbf{k}}}
 - \frac{N_{\mathbf{p}+\mathbf{q},\mathbf{p}+\mathbf{Q}} - N_{u}}
        {-i\Omega_s+u-\epsilon_{\mathbf{p}+\mathbf{q},\mathbf{p}+\mathbf{Q}}}
\right] - \nonumber \\
\frac{\delta n_{\mathbf{k}+\mathbf{Q},\mathbf{k}+\mathbf{q}}
\delta n_{\mathbf{p},\mathbf{p}+\mathbf{q}}}
{i\Omega_s+\epsilon_{\mathbf{k}+\mathbf{q},\mathbf{k}+\mathbf{Q}} - \epsilon_{\mathbf{p}+\mathbf{q},\mathbf{p}}}
&\times&\left[
   \frac{N_{\mathbf{k}+\mathbf{q},\mathbf{k}+\mathbf{Q}} - N_{u}}
   {-i\Omega_s+u-\epsilon_{\mathbf{k}+\mathbf{q},\mathbf{k}+\mathbf{Q}}}
 - \frac{N_{\mathbf{p}+\mathbf{q},\mathbf{p}} - N_{u}}
   {u-\epsilon_{\mathbf{p}+\mathbf{q},\mathbf{p}}}
\right] \bigg\}.
\eea

For diagram $(b)$, $S_2$ is given by the product of
$[ (i\Omega_s-\epsilon_{\mathbf{k}+\mathbf{Q},\mathbf{k}})
(-i\Omega_s-\epsilon_{\mathbf{p}+\mathbf{Q},\mathbf{p}}) ]^{-1}$ and
\bea
\bigg\{
\frac{\delta n_{\mathbf{k},\mathbf{k}+\mathbf{q}} \delta n_{\mathbf{p},\mathbf{p}+\mathbf{q}}}
{\epsilon_{\mathbf{k}+\mathbf{q},\mathbf{k}} + \epsilon_{\mathbf{p}+\mathbf{q},\mathbf{p}}}
&\times&\left[
   \frac{N_u-N_{\mathbf{k}+\mathbf{q},\mathbf{k}}}
        {u-\epsilon_{\mathbf{k}+\mathbf{q},\mathbf{k}}}
 - \frac{1 + N_u + N_{\mathbf{p}+\mathbf{q},\mathbf{p}}}
        {u+\epsilon_{\mathbf{p}+\mathbf{q},\mathbf{p}}}
\right] + \nonumber \\
\frac{\delta n_{\mathbf{k}+\mathbf{Q},\mathbf{k}+\mathbf{q}}
\delta n_{\mathbf{p}+\mathbf{Q},\mathbf{p}+\mathbf{q}}}
{\epsilon_{\mathbf{k}+\mathbf{q},\mathbf{k}+\mathbf{Q}} + \epsilon_{\mathbf{p}+\mathbf{q},\mathbf{p}+\mathbf{Q}}}
&\times&\left[
   \frac{N_u-N_{\mathbf{k}+\mathbf{q},\mathbf{k}+\mathbf{Q}}}
   {-i\Omega_s+u-\epsilon_{\mathbf{k}+\mathbf{q},\mathbf{k}+\mathbf{Q}}}
 - \frac{1+N_u+N_{\mathbf{p}+\mathbf{q},\mathbf{p}+\mathbf{Q}}}
   {-i\Omega_s+u+\epsilon_{\mathbf{p}+\mathbf{q},\mathbf{p}+\mathbf{Q}}}
\right] - \nonumber \\
\frac{\delta n_{\mathbf{k},\mathbf{k}+\mathbf{q}}
\delta n_{\mathbf{p}+\mathbf{Q},\mathbf{p}+\mathbf{q}}}
{-i\Omega_s+\epsilon_{\mathbf{k}+\mathbf{q},\mathbf{k}} + \epsilon_{\mathbf{p}+\mathbf{q},\mathbf{p}+\mathbf{Q}}}
&\times&\left[
   \frac{N_u-N_{\mathbf{k}+\mathbf{q},\mathbf{k}}}
        {u-\epsilon_{\mathbf{k}+\mathbf{q},\mathbf{k}}}
 - \frac{1+N_u+N_{\mathbf{p}+\mathbf{q},\mathbf{p}+\mathbf{Q}}}
        {-i\Omega_s+u+\epsilon_{\mathbf{p}+\mathbf{q},\mathbf{p}+\mathbf{Q}}}
\right] - \nonumber \\
\frac{\delta n_{\mathbf{k}+\mathbf{Q},\mathbf{k}+\mathbf{q}}
\delta n_{\mathbf{p},\mathbf{p}+\mathbf{q}}}
{i\Omega_s+\epsilon_{\mathbf{k}+\mathbf{q},\mathbf{k}+\mathbf{Q}} + \epsilon_{\mathbf{p}+\mathbf{q},\mathbf{p}}}
&\times&\left[
   \frac{N_u-N_{\mathbf{k}+\mathbf{q},\mathbf{k}+\mathbf{Q}}}
   {-i\Omega_s+u-\epsilon_{\mathbf{k}+\mathbf{q},\mathbf{k}+\mathbf{Q}}}
 - \frac{1+N_u+N_{\mathbf{p}+\mathbf{q},\mathbf{p}}}
   {u+\epsilon_{\mathbf{p}+\mathbf{q},\mathbf{p}}}
\right] \bigg\}.
\eea

\vspace{5mm}

\underline{$S_{3}$}. For diagram $(a)$, $S_3$ is given by the product of
$[ (i\Omega_s-\epsilon_{\mathbf{k}+\mathbf{Q},\mathbf{k}})
(i\Omega_s-\epsilon_{\mathbf{p}+\mathbf{Q},\mathbf{p}}) ]^{-1}$ and
\bea
\bigg\{
\frac{\delta n_{\mathbf{k},\mathbf{k}+\mathbf{q}} \delta n_{\mathbf{p},\mathbf{p}+\mathbf{q}}}
{\epsilon_{\mathbf{k}+\mathbf{q},\mathbf{k}} - \epsilon_{\mathbf{p}+\mathbf{q},\mathbf{p}}}
&\times&\left[
   \frac{1+N_u+N_{\mathbf{k}+\mathbf{q},\mathbf{k}}}
        {-i\Omega_s+u+\epsilon_{\mathbf{k}+\mathbf{q},\mathbf{k}}}
 - \frac{1+N_u+N_{\mathbf{p}+\mathbf{q},\mathbf{p}}}
        {-i\Omega_s+u+\epsilon_{\mathbf{p}+\mathbf{q},\mathbf{p}}}
\right] + \nonumber \\
\frac{\delta n_{\mathbf{k}+\mathbf{Q},\mathbf{k}+\mathbf{q}}
\delta n_{\mathbf{p}+\mathbf{Q},\mathbf{p}+\mathbf{q}}}
{\epsilon_{\mathbf{k}+\mathbf{q},\mathbf{k}+\mathbf{Q}} - \epsilon_{\mathbf{p}+\mathbf{q},\mathbf{p}+\mathbf{Q}}}
&\times&\left[
   \frac{1+N_u+N_{\mathbf{k}+\mathbf{q},\mathbf{k}+\mathbf{Q}}}
   {u+\epsilon_{\mathbf{k}+\mathbf{q},\mathbf{k}+\mathbf{Q}}}
 - \frac{1+N_u+N_{\mathbf{p}+\mathbf{q},\mathbf{p}+\mathbf{Q}}}
   {u+\epsilon_{\mathbf{p}+\mathbf{q},\mathbf{p}+\mathbf{Q}}}
\right] - \nonumber \\
\frac{\delta n_{\mathbf{k},\mathbf{k}+\mathbf{q}}
\delta n_{\mathbf{p}+\mathbf{Q},\mathbf{p}+\mathbf{q}}}
{-i\Omega_s+\epsilon_{\mathbf{k}+\mathbf{q},\mathbf{k}} - \epsilon_{\mathbf{p}+\mathbf{q},\mathbf{p}+\mathbf{Q}}}
&\times&\left[
   \frac{1+N_u+N_{\mathbf{k}+\mathbf{q},\mathbf{k}}}
        {-i\Omega_s+u+\epsilon_{\mathbf{k}+\mathbf{q},\mathbf{k}}}
 - \frac{1+N_u+N_{\mathbf{p}+\mathbf{q},\mathbf{p}+\mathbf{Q}}}
        {u+\epsilon_{\mathbf{p}+\mathbf{q},\mathbf{p}+\mathbf{Q}}}
\right] - \nonumber \\
\frac{\delta n_{\mathbf{k}+\mathbf{Q},\mathbf{k}+\mathbf{q}}
\delta n_{\mathbf{p},\mathbf{p}+\mathbf{q}}}
{i\Omega_s+\epsilon_{\mathbf{k}+\mathbf{q},\mathbf{k}+\mathbf{Q}} - \epsilon_{\mathbf{p}+\mathbf{q},\mathbf{p}}}
&\times&\left[
   \frac{1+N_u+N_{\mathbf{k}+\mathbf{q},\mathbf{k}+\mathbf{Q}}}
   {u+\epsilon_{\mathbf{k}+\mathbf{q},\mathbf{k}+\mathbf{Q}}}
 - \frac{1+N_u+N_{\mathbf{p}+\mathbf{q},\mathbf{p}}}
   {-i\Omega_s+u+\epsilon_{\mathbf{p}+\mathbf{q},\mathbf{p}}}
\right] \bigg\}.
\eea

For diagram $(b)$, $S_3$ is given by the product of
$[ (i\Omega_s-\epsilon_{\mathbf{k}+\mathbf{Q},\mathbf{k}})
(-i\Omega_s-\epsilon_{\mathbf{p}+\mathbf{Q},\mathbf{p}}) ]^{-1}$ and
\bea
\bigg\{
\frac{\delta n_{\mathbf{k},\mathbf{k}+\mathbf{q}} \delta n_{\mathbf{p},\mathbf{p}+\mathbf{q}}}
{\epsilon_{\mathbf{k}+\mathbf{q},\mathbf{k}} + \epsilon_{\mathbf{p}+\mathbf{q},\mathbf{p}}}
&\times&\left[
 -  \frac{1+N_u+N_{\mathbf{k}+\mathbf{q},\mathbf{k}}}
        {-i\Omega_s+u+\epsilon_{\mathbf{k}+\mathbf{q},\mathbf{k}}}
 + \frac{N_u-N_{\mathbf{p}+\mathbf{q},\mathbf{p}}}
        {-i\Omega_s+u-\epsilon_{\mathbf{p}+\mathbf{q},\mathbf{p}}}
\right] + \nonumber \\
\frac{\delta n_{\mathbf{k}+\mathbf{Q},\mathbf{k}+\mathbf{q}}
\delta n_{\mathbf{p}+\mathbf{Q},\mathbf{p}+\mathbf{q}}}
{\epsilon_{\mathbf{k}+\mathbf{q},\mathbf{k}+\mathbf{Q}} + \epsilon_{\mathbf{p}+\mathbf{q},\mathbf{p}+\mathbf{Q}}}
&\times&\left[
 -  \frac{1+N_u+N_{\mathbf{k}+\mathbf{q},\mathbf{k}+\mathbf{Q}}}
   {u+\epsilon_{\mathbf{k}+\mathbf{q},\mathbf{k}+\mathbf{Q}}}
 + \frac{N_u-N_{\mathbf{p}+\mathbf{q},\mathbf{p}+\mathbf{Q}}}
   {u-\epsilon_{\mathbf{p}+\mathbf{q},\mathbf{p}+\mathbf{Q}}}
\right] - \nonumber \\
\frac{\delta n_{\mathbf{k},\mathbf{k}+\mathbf{q}} \delta n_{\mathbf{p}+\mathbf{Q},\mathbf{p}+\mathbf{q}}}
{-i\Omega_s+\epsilon_{\mathbf{k}+\mathbf{q},\mathbf{k}} + \epsilon_{\mathbf{p}+\mathbf{q},\mathbf{p}+\mathbf{Q}}}
&\times&\left[
 -  \frac{1+N_u+N_{\mathbf{k}+\mathbf{q},\mathbf{k}}}
        {-i\Omega_s+u+\epsilon_{\mathbf{k}+\mathbf{q},\mathbf{k}}}
 + \frac{N_u-N_{\mathbf{p}+\mathbf{q},\mathbf{p}+\mathbf{Q}}}
        {u-\epsilon_{\mathbf{p}+\mathbf{q},\mathbf{p}+\mathbf{Q}}}
\right] - \nonumber \\
\frac{\delta n_{\mathbf{k}+\mathbf{Q},\mathbf{k}+\mathbf{q}}
\delta n_{\mathbf{p},\mathbf{p}+\mathbf{q}}}
{i\Omega_s+\epsilon_{\mathbf{k}+\mathbf{q},\mathbf{k}+\mathbf{Q}} + \epsilon_{\mathbf{p}+\mathbf{q},\mathbf{p}}}
&\times&\left[
 -  \frac{1+N_u+N_{\mathbf{k}+\mathbf{q},\mathbf{k}+\mathbf{Q}}}
   {u+\epsilon_{\mathbf{k}+\mathbf{q},\mathbf{k}+\mathbf{Q}}}
 + \frac{N_u-N_{\mathbf{p}+\mathbf{q},\mathbf{p}}}
   {-i\Omega_s+u-\epsilon_{\mathbf{p}+\mathbf{q},\mathbf{p}}}
\right] \bigg\}.
\eea

\vspace{5mm}

\underline{$S_{4}$}. For diagram $(a)$, $S_4$ is given by the product of
$[ (i\Omega_s-\epsilon_{\mathbf{k}+\mathbf{Q},\mathbf{k}})
(i\Omega_s-\epsilon_{\mathbf{p}+\mathbf{Q},\mathbf{p}}) (i\Omega_s-u-v)]^{-1} $ and
\bea
&&\bigg\{
\frac{\delta n_{\mathbf{k},\mathbf{k}+\mathbf{q}} \delta n_{\mathbf{p},\mathbf{p}+\mathbf{q}}}
{\epsilon_{\mathbf{k}+\mathbf{q},\mathbf{k}} - \epsilon_{\mathbf{p}+\mathbf{q},\mathbf{p}}}
\times \nonumber \\
&&\;\;\;\;\;\;\;\;\;\;\;\;\;\;\;\;\;\;\;\;\left[
   \frac{1+N_u+N_{\mathbf{k}+\mathbf{q},\mathbf{k}}}
        {-i\Omega_s+u+\epsilon_{\mathbf{k}+\mathbf{q},\mathbf{k}}}
 - \frac{1+N_u+N_{\mathbf{p}+\mathbf{q},\mathbf{p}}}
        {-i\Omega_s+u+\epsilon_{\mathbf{p}+\mathbf{q},\mathbf{p}}}
 + \frac{N_{\mathbf{k}+\mathbf{q},\mathbf{k}}-N_v}
        {v-\epsilon_{\mathbf{k}+\mathbf{q},\mathbf{k}}}
 - \frac{N_{\mathbf{p}+\mathbf{q},\mathbf{p}}-N_v}
        {v-\epsilon_{\mathbf{p}+\mathbf{q},\mathbf{p}}}
    \right] + \nonumber \\
&&\frac{\delta n_{\mathbf{k}+\mathbf{Q},\mathbf{k}+\mathbf{q}} \delta n_{\mathbf{p}+\mathbf{Q},\mathbf{p}+\mathbf{q}}}
{\epsilon_{\mathbf{k}+\mathbf{q},\mathbf{k}+\mathbf{Q}} - \epsilon_{\mathbf{p}+\mathbf{q},\mathbf{p}+\mathbf{Q}}}
\times \nonumber \\
&&\;\;\;\;\;\;\;\;\;\;\;\;\;\;\;\;\;\;\;\;\left[
   \frac{1+N_u+N_{\mathbf{k}+\mathbf{q},\mathbf{k}+\mathbf{Q}}}
   {u+\epsilon_{\mathbf{k}+\mathbf{q},\mathbf{k}+\mathbf{Q}}}
 - \frac{1+N_u+N_{\mathbf{p}+\mathbf{q},\mathbf{p}+\mathbf{Q}}}
   {u+\epsilon_{\mathbf{p}+\mathbf{q},\mathbf{p}+\mathbf{Q}}}
 + \frac{N_{\mathbf{k}+\mathbf{q},\mathbf{k}+\mathbf{Q}}-N_v}
   {-i\Omega_s+v-\epsilon_{\mathbf{k}+\mathbf{q},\mathbf{k}+\mathbf{Q}}}
 - \frac{N_{\mathbf{p}+\mathbf{q},\mathbf{p}+\mathbf{Q}}-N_v}
   {-i\Omega_s+v-\epsilon_{\mathbf{p}+\mathbf{q},\mathbf{p}+\mathbf{Q}}}
   \right] - \nonumber \\
&&\frac{\delta n_{\mathbf{k},\mathbf{k}+\mathbf{q}}
\delta n_{\mathbf{p}+\mathbf{Q},\mathbf{p}+\mathbf{q}}}
{-i\Omega_s+\epsilon_{\mathbf{k}+\mathbf{q},\mathbf{k}} - \epsilon_{\mathbf{p}+\mathbf{q},\mathbf{p}+\mathbf{Q}}}
\times \nonumber \\
&&\;\;\;\;\;\;\;\;\;\;\;\;\;\;\;\;\;\;\;\;\left[
   \frac{1+N_u+N_{\mathbf{k}+\mathbf{q},\mathbf{k}}}
        {-i\Omega_s+u+\epsilon_{\mathbf{k}+\mathbf{q},\mathbf{k}}}
 - \frac{1+N_u+N_{\mathbf{p}+\mathbf{q},\mathbf{p}+\mathbf{Q}}}
        {u+\epsilon_{\mathbf{p}+\mathbf{q},\mathbf{p}+\mathbf{Q}}}
 + \frac{N_{\mathbf{k}+\mathbf{q},\mathbf{k}}-N_v}
        {v-\epsilon_{\mathbf{k}+\mathbf{q},\mathbf{k}}}
 - \frac{N_{\mathbf{p}+\mathbf{q},\mathbf{p}+\mathbf{Q}}-N_v}
        {-i\Omega_s+v-\epsilon_{\mathbf{p}+\mathbf{q},\mathbf{p}+\mathbf{Q}}}
   \right] - \nonumber \\
&&\frac{\delta n_{\mathbf{k}+\mathbf{Q},\mathbf{k}+\mathbf{q}}
\delta n_{\mathbf{p},\mathbf{p}+\mathbf{q}}}
{i\Omega_s+\epsilon_{\mathbf{k}+\mathbf{q},\mathbf{k}+\mathbf{Q}} - \epsilon_{\mathbf{p}+\mathbf{q},\mathbf{p}}}
\times \nonumber \\
&&\;\;\;\;\;\;\;\;\;\;\;\;\;\;\;\;\;\;\;\;\left[
   \frac{1+N_u+N_{\mathbf{k}+\mathbf{q},\mathbf{k}+\mathbf{Q}}}
   {u+\epsilon_{\mathbf{k}+\mathbf{q},\mathbf{k}+\mathbf{Q}}}
 - \frac{1+N_u+N_{\mathbf{p}+\mathbf{q},\mathbf{p}}}
   {-i\Omega_s+u+\epsilon_{\mathbf{p}+\mathbf{q},\mathbf{p}}}
+  \frac{N_{\mathbf{k}+\mathbf{q},\mathbf{k}+\mathbf{Q}}-N_v}
   {-i\Omega_s+v-\epsilon_{\mathbf{k}+\mathbf{q},\mathbf{k}+\mathbf{Q}}}
 - \frac{N_{\mathbf{p}+\mathbf{q},\mathbf{p}}-N_v}
   {v-\epsilon_{\mathbf{p}+\mathbf{q},\mathbf{p}}}
\right] \bigg\}.
\eea

For diagram $(b)$, $S_4$ is given by the product of
$[ (i\Omega_s-\epsilon_{\mathbf{k}+\mathbf{Q},\mathbf{k}})
(-i\Omega_s-\epsilon_{\mathbf{p}+\mathbf{Q},\mathbf{p}}) (i\Omega_s-u-v)]^{-1} $ and
\bea
&&\bigg\{
\frac{\delta n_{\mathbf{k},\mathbf{k}+\mathbf{q}} \delta n_{\mathbf{p},\mathbf{p}+\mathbf{q}}}
{\epsilon_{\mathbf{k}+\mathbf{q},\mathbf{k}} + \epsilon_{\mathbf{p}+\mathbf{q},\mathbf{p}}}
\times \nonumber \\
&&\;\;\;\;\;\;\;\;\;\;\;\;\;\;\;\;\;\;\;\;\left[
 -  \frac{1+N_u+N_{\mathbf{k}+\mathbf{q},\mathbf{k}}}
        {-i\Omega_s+u+\epsilon_{\mathbf{k}+\mathbf{q},\mathbf{k}}}
 + \frac{N_u-N_{\mathbf{p}+\mathbf{q},\mathbf{p}}}
        {-i\Omega_s+u-\epsilon_{\mathbf{p}+\mathbf{q},\mathbf{p}}}
 + \frac{N_v-N_{\mathbf{k}+\mathbf{q},\mathbf{k}}}
        {v-\epsilon_{\mathbf{k}+\mathbf{q},\mathbf{k}}}
 - \frac{1+N_v+N_{\mathbf{p}+\mathbf{q},\mathbf{p}}}
        {v+\epsilon_{\mathbf{p}+\mathbf{q},\mathbf{p}}}
    \right] + \nonumber \\
&&\frac{\delta n_{\mathbf{k}+\mathbf{Q},\mathbf{k}+\mathbf{q}}
\delta n_{\mathbf{p}+\mathbf{Q},\mathbf{p}+\mathbf{q}}}
{\epsilon_{\mathbf{k}+\mathbf{q},\mathbf{k}+\mathbf{Q}} + \epsilon_{\mathbf{p}+\mathbf{q},\mathbf{p}+\mathbf{Q}}}
\times \nonumber \\
&&\;\;\;\;\;\;\;\;\;\;\;\;\;\;\;\;\;\;\;\;\left[
 - \frac{1+N_u+N_{\mathbf{k}+\mathbf{q},\mathbf{k}+\mathbf{Q}}}
   {u+\epsilon_{\mathbf{k}+\mathbf{q},\mathbf{k}+\mathbf{Q}}}
 + \frac{N_u-N_{\mathbf{p}+\mathbf{q},\mathbf{p}+\mathbf{Q}}}
   {u-\epsilon_{\mathbf{p}+\mathbf{q},\mathbf{p}+\mathbf{Q}}}
 + \frac{N_v-N_{\mathbf{k}+\mathbf{q},\mathbf{k}+\mathbf{Q}}}
   {-i\Omega_s+v-\epsilon_{\mathbf{k}+\mathbf{q},\mathbf{k}+\mathbf{Q}}}
 - \frac{1+N_v+N_{\mathbf{p}+\mathbf{q},\mathbf{p}+\mathbf{Q}}}
   {-i\Omega_s+v+\epsilon_{\mathbf{p}+\mathbf{q},\mathbf{p}+\mathbf{Q}}}
   \right] - \nonumber \\
&&\frac{\delta n_{\mathbf{k},\mathbf{k}+\mathbf{q}}
\delta n_{\mathbf{p}+\mathbf{Q},\mathbf{p}+\mathbf{q}}}
{-i\Omega_s+\epsilon_{\mathbf{k}+\mathbf{q},\mathbf{k}} + \epsilon_{\mathbf{p}+\mathbf{q},\mathbf{p}+\mathbf{Q}}}
\times \nonumber \\
&&\;\;\;\;\;\;\;\;\;\;\;\;\;\;\;\;\;\;\;\;\left[
 -  \frac{1+N_u+N_{\mathbf{k}+\mathbf{q},\mathbf{k}}}
        {-i\Omega_s+u+\epsilon_{\mathbf{k}+\mathbf{q},\mathbf{k}}}
 + \frac{N_u-N_{\mathbf{p}+\mathbf{q},\mathbf{p}+\mathbf{Q}}}
        {u-\epsilon_{\mathbf{p}+\mathbf{q},\mathbf{p}+\mathbf{Q}}}
 + \frac{N_v-N_{\mathbf{k}+\mathbf{q},\mathbf{k}}}
        {v-\epsilon_{\mathbf{k}+\mathbf{q},\mathbf{k}}}
 - \frac{1+N_v+N_{\mathbf{p}+\mathbf{q},\mathbf{p}+\mathbf{Q}}}
        {-i\Omega_s+v+\epsilon_{\mathbf{p}+\mathbf{q},\mathbf{p}+\mathbf{Q}}}
   \right] - \nonumber \\
&&\frac{\delta n_{\mathbf{k}+\mathbf{Q},\mathbf{k}+\mathbf{q}}
\delta n_{\mathbf{p},\mathbf{p}+\mathbf{q}}}
{i\Omega_s+\epsilon_{\mathbf{k}+\mathbf{q},\mathbf{k}+\mathbf{Q}} + \epsilon_{\mathbf{p}+\mathbf{q},\mathbf{p}}}
\times \nonumber \\
&&\;\;\;\;\;\;\;\;\;\;\;\;\;\;\;\;\;\;\;\;\left[
 - \frac{1+N_u+N_{\mathbf{k}+\mathbf{q},\mathbf{k}+\mathbf{Q}}}
   {u+\epsilon_{\mathbf{k}+\mathbf{q},\mathbf{k}+\mathbf{Q}}}
 + \frac{N_u-N_{\mathbf{p}+\mathbf{q},\mathbf{p}}}
   {-i\Omega_s+u-\epsilon_{\mathbf{p}+\mathbf{q},\mathbf{p}}}
 + \frac{N_v-N_{\mathbf{k}+\mathbf{q},\mathbf{k}+\mathbf{Q}}}
   {-i\Omega_s+v-\epsilon_{\mathbf{k}+\mathbf{q},\mathbf{k}+\mathbf{Q}}}
 - \frac{1+N_v+N_{\mathbf{p}+\mathbf{q},\mathbf{p}}}
   {v+\epsilon_{\mathbf{p}+\mathbf{q},\mathbf{p}}}
\right] \bigg\}.
\eea

\begin{figure}[tbh]
\centerline{\includegraphics[scale=0.4]{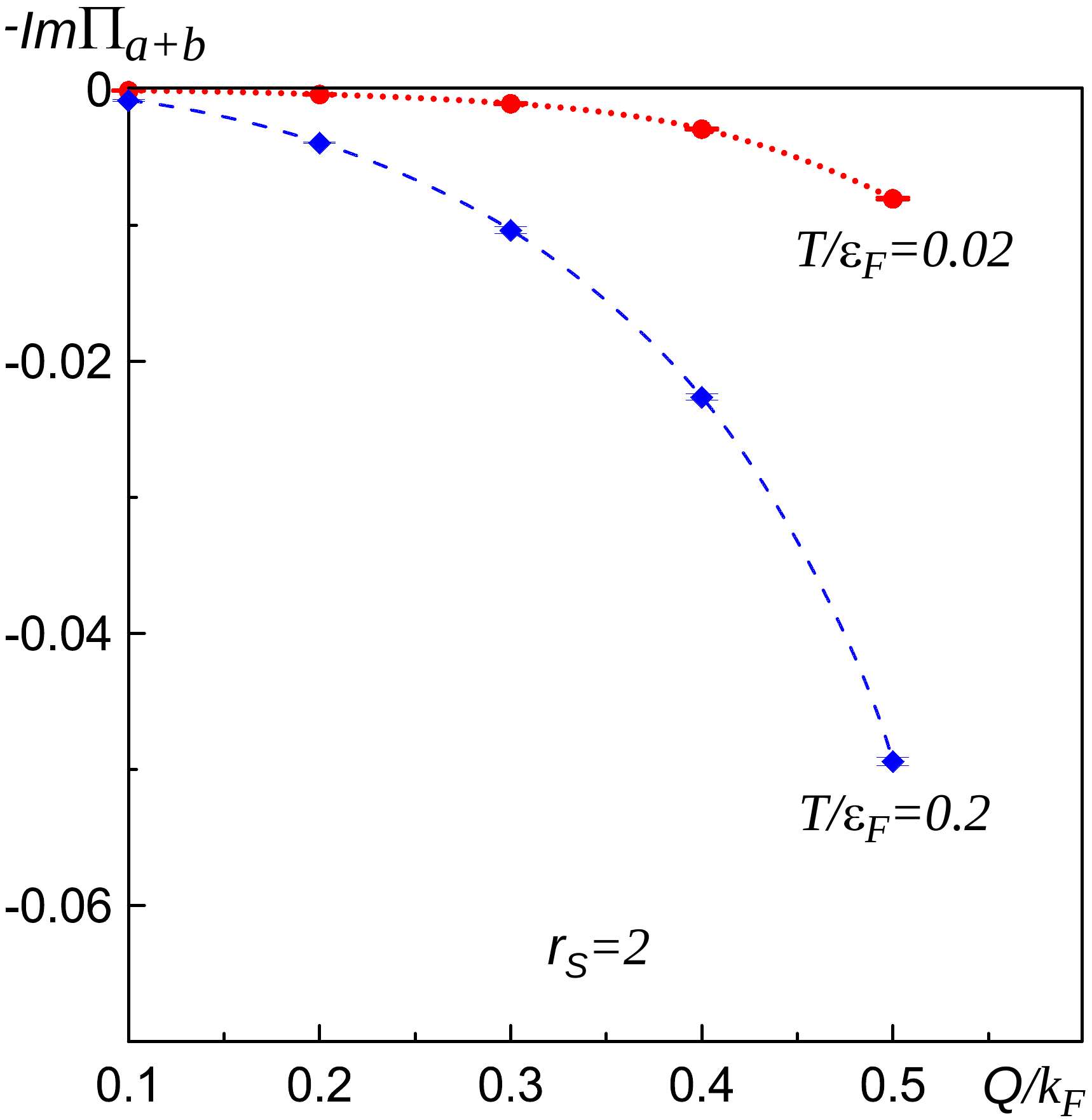}}
\caption{(color online) $-Im\Pi_{a+b}$ as a function of momentum $Q$
for $r_s=2$ and $T/\varepsilon_F=0.02$ and $0.2$. }
\label{Fig7}
\end{figure}

To move to real frequencies, replace $i\Omega_s$ with $\Omega+i\eta$ {\it before} all integrations.
In Fig.~\ref{Fig7} we show our results for $-Im\Pi^{(2)}_{a+b}(\mathbf{Q},\Omega=\omega_{pl}(\mathbf{Q}))$
for $r_s=2$ for $T/\varepsilon_F=0.02$ and $0.2$.
All by itself, this partial contribution is not physical because it has a wrong sign
(the same was true for lower-order vertex corrections).
